\documentclass[aps,prb,twocolumn,showpacs,superscriptaddress,groupedaddress,floatfix]{revtex4-2}
\usepackage{color}
\usepackage{mathtools}
\usepackage{latexsym}
\usepackage{graphicx}
\usepackage{float}
\usepackage{dcolumn}
\usepackage{bm}
\usepackage{amssymb}
\usepackage{amsmath}
\usepackage{mathtools}
\usepackage[space]{grffile}
\usepackage{xcolor}

\newcommand{\be}{\begin{equation}}
\newcommand{\ee}{\end{equation}}
\newcommand{\bea}{\begin{eqnarray}}
\newcommand{\eea}{\end{eqnarray}}

\newcommand{\sx}{\sigma_1^x}
\newcommand{\sy}{\sigma_1^y}

\newcommand{\ainf}{A_\infty}

\begin{document}


\title{Strong and almost strong modes of Floquet spin chains in Krylov subspaces}

\author{Daniel J. Yates}
\author{Aditi Mitra}
\affiliation{Center for Quantum Phenomena, Department of Physics, New York University, 726 Broadway, New York, NY, 10003, USA}

\date{\today}

\begin{abstract}
  Integrable Floquet spin chains are known to host
  strong zero and $\pi$ modes which are boundary operators that respectively commute and
  anticommute with the Floquet unitary generating stroboscopic time-evolution,
  in addition to anticommuting with a discrete symmetry of the Floquet unitary. Thus the existence of strong modes
  imply a characteristic pairing structure of the full spectrum. Weak interactions modify the strong modes
  to almost strong modes that almost commute or anticommute with the Floquet unitary. Manifestations
  of strong and almost strong modes are presented in two different Krylov subspaces.
  One is a Krylov subspace obtained from a Lanczos iteration that maps the
  time-evolution generated by the Floquet Hamiltonian onto dynamics of a single particle on
  a fictitious chain with nearest neighbor hopping.
  The second is a Krylov subspace obtained from the Arnoldi iteration that maps the time-evolution generated
  directly by the Floquet unitary onto dynamics of a single particle on a fictitious chain with longer range hopping.
  While the former Krylov subspace is
  sensitive to the branch of the logarithm of the Floquet unitary, the latter obtained from the Arnoldi  scheme is not. The effective single particle
  models in the Krylov subspace are discussed, and the topological properties of the Krylov chain that
  ensure stable $0$ and $\pi$ modes at the boundaries are highlighted. The role of interactions is discussed.
  Expressions for the lifetime of the
  almost strong modes are derived in terms of the parameters of the Krylov subspace, and are compared with exact diagonalization.
\end{abstract}
\maketitle

\section{Introduction} \label{Intro}

The Kitaev chain \cite{Kitaev01}, which after a Jordan-Wigner transformation maps to the transverse field Ising model
(TFIM) \cite{Sachdevbook}, despite its apparent simplicity, is a toy model for understanding
diverse phenomena such as quantum phase transitions \cite{Sachdevbook} and topological systems \cite{Bernevigbook}. It also forms
a building block for realizing non-Abelian braiding \cite{Kitaev06,NayakRMP08,Alicea12,Freedman15}.
In recent years, periodic or Floquet driving of this model has helped conceptualize new
phenomena such as boundary and bulk discrete time crystals \cite{Sacha2018,Else2019,Khemani2019}.

A key feature of the TFIM model, and its anisotropic generalization,
is that it hosts robust edge modes known as strong zero modes (SZMs) \cite{Kitaev01,Fendley16}.
An operator $\Psi_0$ is a SZM if it obeys certain properties. Firstly it  should
commute with the Hamiltonian $H$ in the thermodynamic limit i.e,  $\left[\Psi_0,H\right]\approx 0$,
secondly it should anticommute with
a discrete (say ${\mathbb Z}_2$) symmetry
of the Hamiltonian ${\mathcal D}$, $\{\Psi_0,{\mathcal D} \}=0$,
and thirdly it should be a local operator with the property $\Psi_0^2 = O(1)$. The presence of a SZM immediately implies that the
entire spectrum of $H$ is doubly degenerate, where the degenerate pairs are $\{|n\rangle, \Psi_0|n\rangle\}$,
with each member of the pair being eigenstates of ${\mathcal D}$, but in the opposite symmetry sector. This eigenstate-phase
makes the edge modes extremely stable to adding symmetry preserving perturbations, such as exchange interactions between spins.
In particular, the edge modes
acquire a finite lifetime in the presence of interactions,
but with the lifetime being non-perturbative in the strength of the interactions \cite{Nayak17,Fendley17,Parker19,Laumann20,Yates20,Yates20a}.
These long-lived quasi-stable
modes in the presence of perturbations are referred to as almost strong zero modes (ASZMs) \cite{Fendley17}.

Under Floquet driving, besides the SZMs, new edge modes arise \cite{Zoller11,Brandes12,Sen13, Delplace14,Yates19},
which are called strong $\pi$ modes (SPMs) \cite{Yates19},
with the
possibility of having phases where SZMs and SPMs co-exist \cite{Khemani16,Kyser16,Kyser-I, Kyser-II,Yates19}.
In order to define the strong modes in a Floquet system,
one needs to adapt the previous definition of the SZM to that of a Floquet unitary over one drive cycle, $U$.
The SZM $\Psi_0$ and the SPM $\Psi_{\pi}$ are now defined as follows.
Both these operators anticommute with the discrete symmetry of $U$, $\{\Psi_0,{\mathcal D}\}=0, \{\Psi_{\pi}, {\mathcal D}\}=0$.
But while the SZM $\Psi_0$
commutes with the Floquet unitary in the thermodynamic limit $\left[\Psi_0,U\right]\approx 0$, the SPM $\Psi_{\pi}$
anticommutes with it $\{\Psi_\pi,U\}\approx 0$ in the same limit. Moreover, as for the static case,
both operators are local, with the property $\Psi_{0,\pi}^2 = O(1)$.
Thus existence of $\Psi_0$ implies that the entire spectrum of $U$ is doubly degenerate $\{|n\rangle, \Psi_0|n\rangle\}$,
with the two eigenstates of a pair also being  eigenstates of opposite symmetry. The existence of $\Psi_\pi$ also implies
that the spectrum is paired, but with each pair $|n\rangle$, $\Psi_{\pi}|n\rangle$ not only being
eigenstates of opposite symmetry, but also being separated by the quasi-energy
of $\pi/T$, with $T$ being the period of the drive. In particular, the eigenstate pair $|n\rangle, \Psi_{\pi}|n\rangle$
pick up a relative phase of $-1$ after odd stroboscopic time-periods. 
Thus the SPM is an example of a boundary time-crystal.

The Majorana mode of the Kitaev chain is an example of a SZM discussed above, but written in the Majorana basis.
  From the discussion above it follows that 
  with open boundary conditions, not just the ground state,
  but all the eigenstates of the Kitaev chain are doubly degenerate in the topologically non-trivial phase. In a similar manner, Floquet driving
  the Kitaev chain gives rise to $\pi$ Majorana modes. These are examples of the SPMs discussed above, with their existence implying a
  $\pi$-pairing for the entire spectrum of the Floquet unitary.

  The strong modes are a property of the entire Hilbert space. When interactions are included in the Kitaev chain, the SMs reduce to ASMs,
  i.e, the exact pairing structure of the entire Hilbert space reduces to an approximate pairing. However even with interactions, there is 
an integrable limit, corresponding to the XYZ chain after a Jordan-Wigner transformation, where SMs survive \cite{Fendley16}.
For a generic non-integrable chain which does not have SMs, when the interactions are weak and hence irrelevant in a renormalization group sense,
one can still have zero modes that ensure a degeneracy only in the
ground state sector. These zero modes are known as weak zero modes \cite{Hosho15, Hosho18, Tohyama21}.

In the Floquet setting, SZMs and SPMs have been identified in free fermion (integrable) Floquet
chains \cite{Zoller11,Sen13,Yates19}. In addition,
the effect of weak interactions was explored using exact diagonalization (ED) where
it was found that quasi-stable and long lived edge modes exist despite bulk heating \cite{Yates19}.  Drawing inspiration from
the static case \cite{Fendley17}, these quasi-stable modes are called
ASZMs and almost strong $\pi$ modes (ASPMs) \cite{Yates19}. The current paper
proposes a route to understanding these quasi-stable modes by mapping their dynamics to a Krylov subspace. 
A similar study was performed for ASZMs in static chains \cite{Yates20,Yates20a}. In addition, a recent paper
showed how the Krylov method can be generalized to Floquet systems, where the method used was a Lanczos iteration
scheme based on the Floquet Hamiltonian \cite{Yates21a}.
Here we further expand on this method, and also introduce the Arnoldi method, which is based on the Floquet unitary rather
than the Floquet Hamiltonian, as a promising alternate route.
Since the Krylov subspace dynamics, whether generated by the Lanczos method or the Arnoldi method, is the dynamics of
a free particle on a fictitous one-dimensional (1$d$) lattice, the Krylov chain, this mapping paves the way for developing analytic
methods to extract the lifetimes for general interacting and driven settings.

The paper is organized as follows. The integrable model is introduced in Section \ref{Model}, and its phases outlined.
In Section \ref{Krylov} the Heisenberg time-evolution of the SZM and SPM are mapped to a Krylov subspace via
a Lanczos iteration scheme where the generator of the dynamics is
the logarithm of the Floquet unitary over one drive cycle, i.e., the Floquet Hamiltonian.
In Section \ref{Arnoldi}, the dynamics generated by
the Floquet unitary is mapped to a Krylov subspace via
the Arnoldi iteration \cite{Arnoldi51}. In section \ref{Arnoldi-free} analytic expressions of the Krylov subspace arising from the
Arnoldi method are derived in the free limit.
In Section \ref{LFAA}, the Arnoldi iteration is used to arrive
at a compact expression for the lifetime of the edge modes that holds for finite size systems
as well as non-zero interactions.
In Section \ref{Int}, the effect of interactions is discussed, and the lifetime of the edge modes
obtained from Krylov space methods are compared with ED.
We present our conclusions in Section \ref{Conclu}.

\begin{figure}
  \includegraphics[width = .49\textwidth]{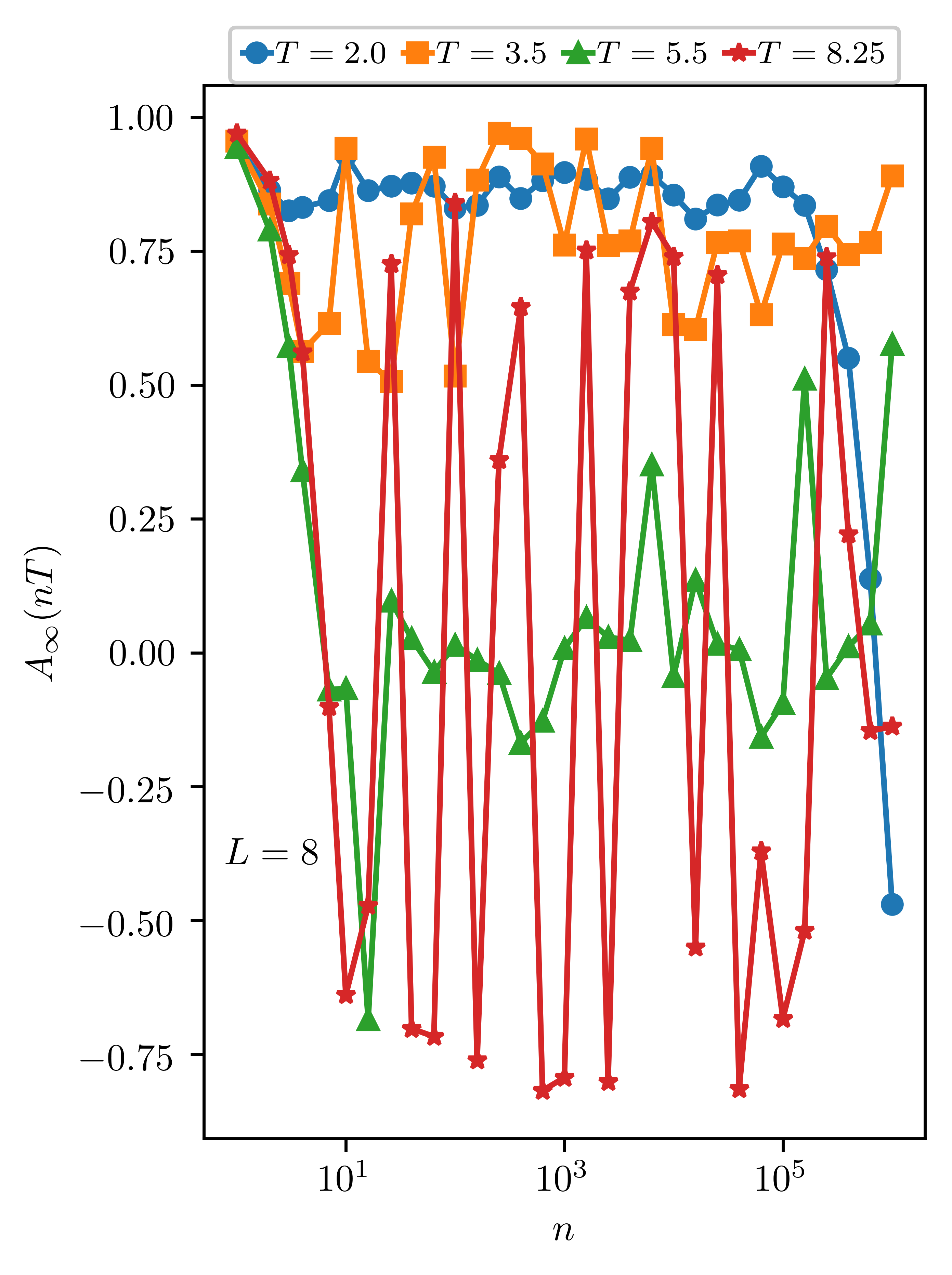}
  \caption{The autocorrelation function $A_{\rm \infty}$ at logarithmically separated stroboscopic times for a
    chain of length $L=8$ and for the free binary drive.
    The parameters are $g=0.3$ and $T=2.0,3.5,5.5,8.25$ corresponding respectively to  SZM phase,
    SZM-SPM phase, trivial phase, and a SPM phase.}
  \label{fig1}
\end{figure}

\section{Model} \label{Model}

In this section we introduce the integrable model
of the Floquet chain. We consider an open chain of length $L$ where
the stroboscopic time evolution is generated
by the following Floquet unitary \cite{Sen13,Khemani16,Kyser-I,Prosen00,Gritsev17,Yates19}
\begin{equation}
  U = e^{-i \frac{T}{2} J_x H_{xx}} e^{-i \frac{T}{2} g H_z}, \label{eqU1}
\end{equation}
where
\begin{align}
  H_z &= \sum_{i=1}^L \sigma^z_i,\\
  H_{xx} & = \sum_{i=1}^{L-1}\sigma^x_i \sigma^x_{i+1}.
\end{align}
In what follows we set $J_x=1$. Eq.~\eqref{eqU1} is a binary drive where the first part of
the drive involves time-evolution purely by the local magnetic field of strength $g T$, while the
second part of the drive involves time-evolution with respect to the nearest-neighbor
Ising interaction of strength $T$.
The Floquet unitary has a discrete symmetry as it commutes with
\begin{align}
   \mathcal{D} = \sigma_1^z \dots \sigma_L^z.\label{eqD}
\end{align}
When we add interactions in Section \ref{Int}, $\mathcal{D}$ will continue to be a symmetry of the problem.

When $T\ll 1$, a high frequency expansion gives the Floquet Hamiltonian
$H_F= H_{xx}/2 + g H_z/2$.  Moreover, in this high frequency limit, $T$ can be interpreted as the period of the drive,
with the stroboscopic time-evolution being $U = e^{-i H_F T}$. 
The phases of the Kitaev chain are recovered in the high frequency limit,
where as $g$ is tuned from $g<1$ to $g>1$, one encounters a quantum phase transition
from a topologically non-trivial phase to a topologically trivial phase.
We are interested in general $T$ where the phase diagram is much richer. Eventually we
are interested in how these phases manifest in the Krylov subspace.

It is convenient to map the problem to Majorana fermions as follows
\begin{align}
  a_{2\ell -1} = \prod_{j=1}^{\ell-1}\sigma^z_j \sigma^x_\ell;\,\, a_{2\ell} = \prod_{j=1}^{\ell-1}\sigma^z_j \sigma^y_\ell,
\end{align}
where $\ell = 1 \ldots L$.
Denoting the vector
\begin{align}
  \vec{a} = \begin{pmatrix} a_1 \\ a_2 \\ a_3 \\\vdots \\a_{2L}\end{pmatrix},
\end{align}
the stroboscopic time-evolution of $\vec{a}$ is as follows
\begin{align}
  U^{\dagger}\vec{a} U = K \vec{a},\label{eqMT}
\end{align}
where on defining
\begin{subequations}\label{eqcsdef}
  \begin{align}
    c_1 &= \cos(Tg),\\
    c_2 &= \cos(T),\\
    s_1 &= \sin(Tg),\\
    s_2 &= \sin(T),
  \end{align}
\end{subequations}
we find that $K$ takes the following form \cite{Yates19} for $L=3$
\begin{align}
   K&=
  \begin{pmatrix}
    c_1& -s_1& 0& 0 &0 &0 \\
    s_1c_2 & c_1 c_2& -c_1s_2& s_1 s_2 &0& 0\\
    s_1 s_2&c_1s_2&c_1c_2 &-s_1c_2 & 0&0\\
    0& 0& s_1 c_2 & c_1c_2 &-c_1 s_2 & s_1 s_2 \\
    0&0 & s_1 s_2 &c_1s_2 & c_1c_2 & -s_1c_2 \\
    0&0 &0 & 0&s_1 & c_1
  \end{pmatrix}.
  \label{eqM}
\end{align}
$K$ is an orthogonal matrix which, depending on the parameters can admit eigenvalues at $\pm 1$.
While $K$ has been shown explicitly for $L=3$, the form for general $L$ can be easily guessed as the
bulk consisting of the second and third rows are repeated. For the example of $L=3$ given above,
we see that the bulk structure is repeated twice.

A key quantity that we will analyze is the autocorrelation function of $\sigma^x_1 =a_1$,
evaluated at stroboscopic times
\begin{subequations}\label{eqAinf}
\begin{align}
  A_{\infty}(n T) &= \frac{1}{2^L}{\rm Tr} \biggl[\sigma^x_1(n T) \sigma^x_1(0)\biggr],\\
  \sigma^x(n T) &= \left[U^{\dagger}\right]^n \sigma^x_1 U^n .
\end{align}
\end{subequations}
When strong modes exist, this
quantity has a non-zero overlap with them, and therefore its stroboscopic
time-evolution is a diagnostic for whether strong modes exist or not. This point will be further clarified below
when we discuss some exactly solvable limits. 
Moreover, from Eq.~\eqref{eqMT}
it is clear that the above autocorrelation function is completely determined
by the action of $K$ on an initial vector that is localized on the first site. Thus $\pm 1$ eigenvalues of
$K$ imply SZMs and SPMs respectively.

We now discuss some exactly solvable limits where $\sigma^x_1$ is the strong mode, or has an $O(1)$ overlap with it.
Below $n$ is an integer.
\begin{itemize}
\item{\(Tg = (2n+1)\pi\), \(T\) arbitrary:} We have \(e^{-iTgH_z/2}
  \propto {\mathcal D}\), thus \(U^\dagger \sx U = \mathcal{D} \sx \mathcal{D} = -\sx,\) and a
  SPM exists. There is however no SZM.

\item{\(Tg = (2n)\pi\), \(T\) arbitrary:} Now \(e^{-iTg H_z/2}
  \propto 1\), thus \(U^\dagger \sx U = \sx\). We now have a SZM while there
  is no SPM.

\item{\(T = (2n+1)\pi\), \(Tg\) arbitrary:} Now we have \(e^{- i
  TH_{xx}/2} \propto \sigma_1^x \sigma_L^x\).  With \(e^{-i T
  g H_z/2} = \prod_{l=1}^L \left[ \cos (T g/2) - i \sigma_l^z \sin(T
  g/2)\right]\). It is straightforward to check that
  \begin{align}
    U^\dagger \sx U &= \cos(Tg) \sx - \sin(Tg) \sy, \nonumber \\
    U^\dagger \sy U &=-\sin(Tg) \sx - \cos(Tg) \sy.\label{eqU3}
  \end{align}
The above linear combinations of $\sigma^x_1$ and $\sigma^y_1$ yield
a SZM and a SPM.

\item{\(T = 2n\pi\), \(Tg \) arbitrary:} We have \(e^{-i
  TH_{xx}/2} \propto 1\). Then,
\begin{align}
    U^\dagger \sx U &= \cos(Tg) \sx - \sin(Tg) \sy, \nonumber \\
    U^\dagger \sy U &=\sin(Tg) \sx + \cos(Tg) \sy. \label{eqU4}
\end{align}
The above linear combinations do not yield  any strong modes unless \(Tg =
  m \pi\) where $m$ is any integer.  Here two cases arise depending
  on whether $m$ is even or odd.
  If \(T g = 2 m \pi\) then \(\sx\) is trivially a SZM because
  \(U \propto 1\).
  If \(Tg = (2 m+1)\pi\), then \(\sx\) is a SPM because \(U\propto {\mathcal D}\).
\end{itemize}
The above examples show that whenever a SPM exists, the Floquet unitary has a characteristic non-local structure by either
being of the form $U\propto {\mathcal D}$ (SPM phase) or being of the form $U \propto \sigma^x_1 \sigma^x_L$ (SZM-SPM phase)
\cite{Khemani16,Kyser16,Kyser-I, Kyser-II}.
We will show later that this non-local structure will have implications on the Krylov subspace.

The above phases are stable to perturbing around it.
Fig.~\ref{fig1} shows
$A_\infty$ for some special cases for stroboscopic and logarithmically separated times.
The parameters chosen are $g=0.3$,
and $T=2.0, 3.5, 5.5, 8.25$. We will work with these parameters for the rest of
the paper when presenting the numerical results.
Among the chosen parameters,
all of them except $T=5.5$ host strong modes. Thus for
all parameters except $T=5.5$, $A_\infty$ is long-lived with its lifetime only set by the system
size $L$. Among these different cases, $T=2.0$ is a phase that has a SZM,
$T=3.5$ is a phase that simultaneously hosts a SZM and a SPM, while $T=8.25$ hosts a SPM.
For a SPM, the stroboscopic time-evolution flips between $O(1)$ positive and negative values,
consistent with period-doubled dynamics.

When both SZM and SPM exist, $\sigma^x_1$ has an overlap with both of them. Thus we can write,
$\Psi_0 = a \sigma^x_1 + \ldots $ and $\Psi_{\pi}=b \sigma_1^x (-1)^n + \ldots$ where $a, b=O(1)$
and positive $a,b>0$.
After the initial transients, we can
write the form of $\sigma^x_1(t)$ as $\sigma^x_1(t) \approx c_0 \Psi_0 + c_\pi \Psi_{\pi} + \ldots$, where $c_{0,\pi}\leq 1/2$
for normalization. Thus the autocorrelation function becomes
$A_{\infty}(n T) \approx c_0 a + c_{\pi} b (-1)^n$. This shows that the signal will oscillate between $c_0 a + c_{\pi} b$
and $c_0 a - c_{\pi} b$. There are several indications that $c_0 > c_{\pi}$. Firstly, note from Fig.~\ref{fig1} that when only a SZM
is present, $c_0\approx 1$, while when only a SPM is present $c_{\pi}\approx 0.75$.
This is further corroborated at
the exactly solvable limit for the SZM-SPM phase discussed
in Eq.~\eqref{eqU3}. Here one finds that at odd stroboscopic times $A_{\infty}(n=1) =\cos(g T)$, while at
even stroboscopic times $A_{\infty}(n=2) = 1$. Since $c_0 > c_{\pi}$,
the signal oscillates between two positive numbers, as shown in Fig.~\ref{fig1}.

Let us define the Floquet Hamiltonian $H_F$ as
\begin{equation}
  H_F = \frac{i}{T} \ln(U). \label{eqHF}
\end{equation}
Note that for the binary drive in Eq.~\eqref{eqU1}, since the problem is free, the matrix $K$ in Eq.~\eqref{eqMT}
and $U$ are simultaneously diagonalized. Thus $i\ln{K}$ directly gives us the Floquet Hamiltonian $H_F$ in the Majorana basis.

The structure of $\ln{K}$ in the vicinity of some exactly solvable points was discussed in Ref.~\onlinecite{Yates21a}.
In particular it was shown that when $T\ll 1$, and $|gT-\pi| \ll 1$, i.e, parameters for
which the system hosts a SPM, $i\ln{K}$ has the form of a topologically non-trivial
Su-Schrieffer-Heeger (SSH) model \cite{SSH79,SSH80}, upto
an overall diagonal matrix $\pi$. Thus the energy of the zero mode of the SSH model is shifted by $\pi$, giving a SPM. Moreover,
such an interpretation survived even when one tuned away from this exactly solvable limit. Ref.~\onlinecite{Yates21a}
also showed that an interpretation of the strong modes in terms of boundary modes of generalized SSH models, also survive in 
the Krylov subspace constructed from the Lanczos method. We will further
expand on this point below, and also discuss the appearance of SSH-type chains in
the Krylov subspace constructed from the Arnoldi method.

\section{Krylov chain from the Lanczos iteration} \label{Krylov}

The time evolution of an operator in a generic integrable or
non-integrable system can be mapped to single particle dynamics on a
semi-infinite chain employing a recursive Lanczos scheme \cite{Recbook}. This method has made a reapparance
recently as a way to identify chaotic dynamics
\cite{Altman19,Gorsky19,Sinha19,Avdoshkin19}.
In this section we outline this
method.  The exponential complexity of solving the dynamics enters
into the calculation of the hopping parameters on this chain which we
denote by $b_n$.

Note the definition of the Floquet Hamiltonian $H_F$ in Eq.~\eqref{eqHF}.
The stroboscopic time-evolution after $m$ periods can be written in terms of $H_F$ as follows
\begin{equation}
  \left[U^{\dagger}\right]^m O U^m = e^{i H_F m T}O e^{-i H_F m T}=\sum_{n = 0}^\infty \frac{(i m T)^n}{n!} \mathcal{L}^n O,
\end{equation}
where we define
\begin{equation}
  \mathcal{L}O = [H_F,O].
\end{equation}
To employ the Lanczos algorithm, we recast the operator dynamics into
vector dynamics by defining $|O) = O$.  Since we are concerned with
infinite temperature quantities, we have an unambiguous choice for an
inner product on the level of the operators,
\begin{equation}
  (A|B) = \frac{1}{2^L} \text{Tr} \left[ A^\dagger B \right].
\end{equation}
The Lanczos algorithm iteratively finds the operator basis that
tri-diagonalizes $\mathcal{L}$. We begin with the seed ``state'',
$|O_1)$, and let $\mathcal{L}|O_1) = b_1 |O_2)$, where $b_1 =
\sqrt{|\mathcal{L}|O_1)|^2}$.  The recursive definition for the basis
operators $|O_{n\ge 2})$ is,
\begin{equation}
  \mathcal{L} |O_n) = b_{n} |O_{n+1}) + b_{n-1} |O_{n-1}),
  \label{eq_lanczos}
\end{equation}
where we define $b_{n\ge 2} = \sqrt{|\mathcal{L}|O_{n})-b_{n-1}|O_{n-1})|^2}$.  It is
straightforward algebra to check that the above procedure will
iteratively find basis operators that yield a $\mathcal{L}$ which is
tri-diagonal, and of the following form,
\begin{equation}\label{Lmat}
  \mathcal{L} =
  \begin{pmatrix}
    & b_1 &     &     & \\
    b_1 &     & b_2 &     & \\
    & b_2 &     & \ddots & \\
    &     & \ddots&   &
  \end{pmatrix}.
\end{equation}
This basis spanned by $|O_n)$ lies within the Krylov subspace of
$\mathcal{L}$ and $|O_1)$. We refer to this tri-diagonal matrix as the
Krylov Hamiltonian $H_K$,
\begin{equation}
  H_K = \sum_n b_n ( c_n^\dagger c_{n+1} + c_{n+1}^\dagger c_n ),\label{eqHK}
\end{equation}
and the 1$d$ lattice it represents, the Krylov chain.

An important aspect of this technique, often overlooked when
discussing chaos, is that the values of $b_n$ are highly dependent on
the choice of seed operator. Further, outside of special cases, namely
a Hamiltonian that is free, the exact solution to the operation
$\mathcal{L}|O_n)$ will require ED, or similar methods with equivalent
costs. This method does not escape the rapidly growing exponential
wall of complexity.  In cases where the calculation of all $b_n$ are
possible, the above algorithm will return a value of
$b_{\text{end}} =0$ for the very last $b_n$. If the iteration is continued further, all the previous
$b_n$s will be repeated until $b_{\rm end}=0$ is again reached.

\begin{figure*}
  \includegraphics[width =0.99\textwidth]{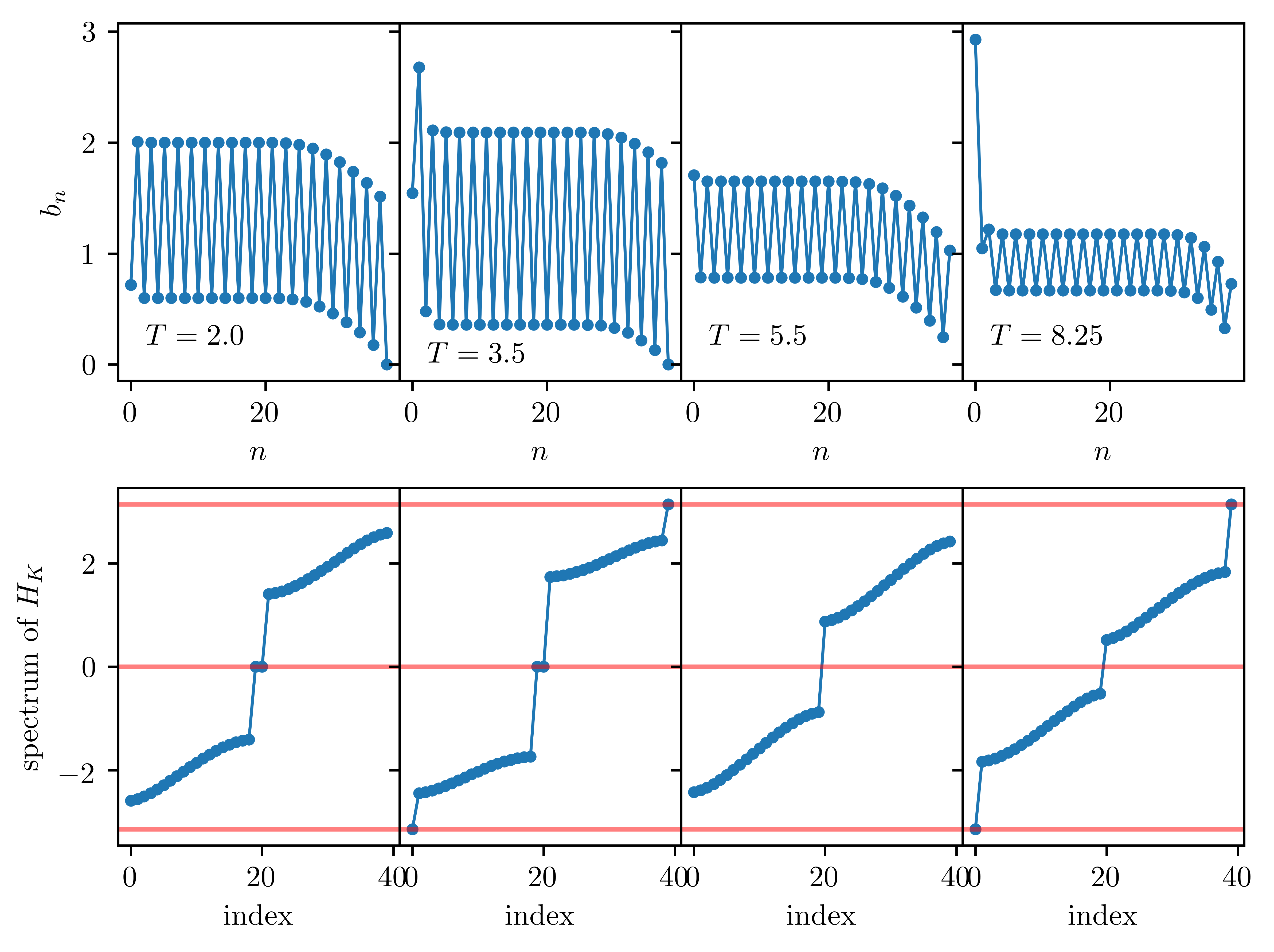}
  \caption{Upper panel: The $b_n$s for the binary drive with $g=0.3$ and system size $L=20$. Lower panel: The corresponding
    spectra with $x$-axis label ``index'' denoting the location of the eigenvalues arranged in ascending order. For both upper and lower panels,
    from left to right $T=2.0,3.5,5.5,8.25$. These parameters correspond to SZM, SZM-SPM, trivial, and SPM phases respectively.
    Horizontal red lines in lower panels correspond to energies $0$ and $\pm \pi$ in units of $T^{-1}$. }
  \label{fig2}
\end{figure*}

For free systems, the operation $\mathcal{L}|O_n)$ can be efficiently
solved when in the Majorana basis. If the starting operator is a
single Majorana then the dimension of the Krylov subspace of that
operator will scale as $2L$, as free system dynamics can only mix the
individual Majoranas among themselves. Outside of free problems,
the size of the full set of $|O_n)$ will be
large. For example, a system size of \(L\) will have $\sim 2^{2L}$
possible basis operators. For all intents and purposes we treat
$\mathcal{L}$ as a semi-infinite chain.  If the number of solved $b_n$
is insufficient for the quantity of interest, an approach that works well is
to supplement the known set with approximate $b_{n}$ that are
calculated based off of trends established among the known hoppings \cite{Yates20,Yates20a}.

Starting with the seed state $|O_1) = |\sigma_1^x)$, we can recast
$A_{\infty}$ into the following form
\begin{equation}
  A_{\infty}(n T) = (e^{i \mathcal{L} n T})_{1,1}.\label{eqAL}
\end{equation}
Now, following the above discussion, the dynamics of $A_{\infty}$ has
been transformed into that of a semi-infinite single-particle problem.
The details of the semi-infinite chain will be discussed in subsequent
sections. Ref.~\onlinecite{Yates20} showed that the slow dynamics of
$A_\infty$ for static spin chains was a result of topological modes residing at the left
boundary (origin) of the Krylov chain.
We discuss the analogous situation for the Floquet problem.

\begin{figure}
  \includegraphics[width =.49\textwidth]{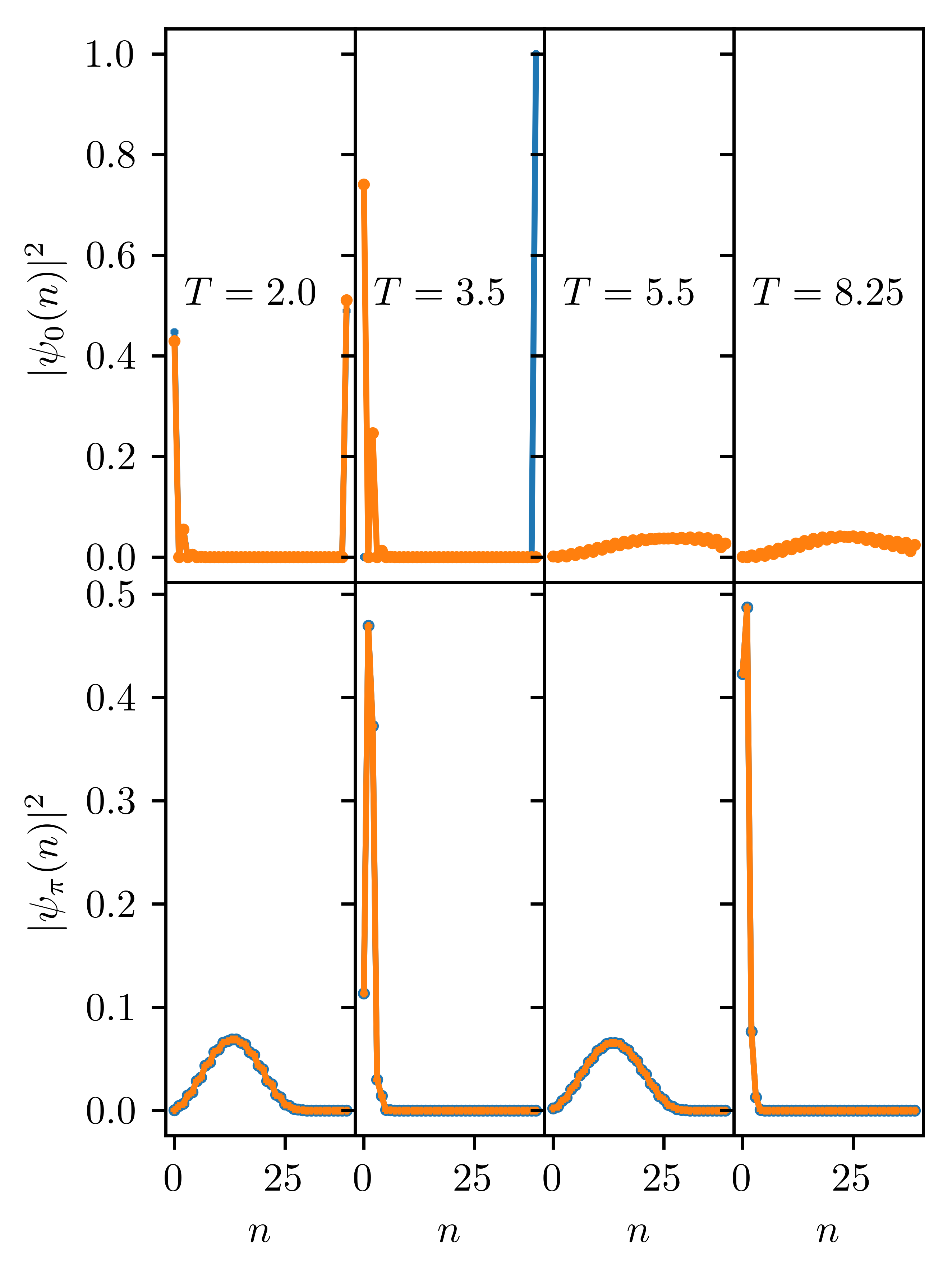}
  \caption{The mod-square of the eigenfunctions of the Krylov chain for the same parameters as Fig.~\ref{fig2}. Upper (lower) panels
    correspond to modes at zero ($\pm \pi$) energy. The orange and blue data sets in each panel reflect the fact that
    these modes appear in pairs.  In particular, there are two zero edge modes and two
    $\pi$ edge modes, the latter occurring at $\pm \pi$ energy. The plots show that $T=5.5$
    hosts no edge modes, consistent with a trivial phase.
    On the other hand $T=2.0$ hosts a pair of edge modes at zero and no edge modes at $\pi$ energy. $T=8.25$
  hosts a pair of edge modes at $\pm\pi$ but no edge modes at $0$ energy. $T=3.5$ hosts edge modes at both $0$ and $\pm\pi$ energies.}
  \label{fig3}
\end{figure}

Since for the binary drive, the problem is free, the matrix $K$ in Eq.~\eqref{eqMT}
and $U$ are simultaneously diagonalized. Thus we construct the Krylov Hamiltonian by taking the Floquet Hamiltonian
to be $ T H_F = i \ln(K)$ and performing the iterative steps outlined above. The parameters $b_n$ of the Krylov
Hamiltonian, and the corresponding spectra are shown in Fig.~\ref{fig2}.
The eigenfunctions with eigenvalues corresponding to $0$ and $\pm \pi$ are
shown in Fig.~\ref{fig3}.

Fig.~\ref{fig2} clearly shows that the Krylov chain has a dimerized structure like that of a SSH chain.
The case of the SZM phase at $T=2.0$ and the trivial phase at $T=5.5$ are the easiest to understand as these
correspond to SSH chains of opposite sign of the dimerization, one sign being topologically trivial ($T=5.5$, with $b_{\rm odd}> b_{\rm even}$)
and the other being topologically non-trivial ($T=2.0$ with  $b_{\rm odd}< b_{\rm even}$).
For the case of the SPM at $T=8.25$ (discussed in detail in Ref.~\onlinecite{Yates21a}),
one finds that the very first hopping is pinned at a large value of $O(\pi)$, while the rest
of the chain, from $n\geq 4$ onwards has a topologically non-trivial dimerization, implying an edge mode. The local strong hopping at the
first site results in an edge mode that is pinned at $\pm \pi$. Thus
$\pi$ modes of the Floquet chain appear as regular zero modes of the topologically non-trivial Krylov chain with an overall shift in energy
by $\pi$ due to boundary conditions imposed by the hoppings on the first few sites.
For the more complicated SZM-SPM phase at $T=3.5$, the first two sites at the edges act as a
two-level system that can host a pair of edge modes. These
modes act as the boundary of an SSH chain with non-trivial sign of the dimerization. This point will be demonstrated further when we discuss
the Krylov Hamiltonian
obtained from the Arnoldi method. Also note that all the eigenvalues of $H_K$ (lower panels of
Fig.~\ref{fig2}) lie within the Floquet Brillouin Zone (FBZ) defined
as $\epsilon T \in \left[-\pi, \pi\right]$. However, this outcome is not guaranteed, and as we will discuss in detail later, a different
basis choice can lead to an unfolded spectrum for the Krylov chain.

Fig.~\ref{fig3} plots the mod-square of modes at $0$ and $\pm\pi$ energies ($|\psi_{0,\pi}(n)|^2$) for all the phases. Note that these modes appear in pairs in
that, when a zero edge mode exists, there are two of them (orange and blue lines in the upper panels of the plot). While when
a $\pi$ mode exists, there are a pair of them located at $\pi$ and $-\pi$ (orange and blue lines in the lower panels of the plot).
The periodicity of the
spectrum under Floquet driving implies that the $\pi$ modes are doubly degenerate in the same way as the $0$ modes are because
$-\pi = \pi + 2\pi n$, where $n$ is any integer.

Fig.~\ref{fig3} shows that the degenerate pairs of edge modes are
not symmetrically localized on the two ends of the Krylov chain as the chain is constructed for the
operator $\sigma^x_1$, and is therefore heavily weighted on the left edge.
If the seed operator would have been a symmetric linear combination of operators on the first and the last site of
the physical chain, then the edge modes of the Krylov chain would also have been symmetrically located at the two
ends of the chain. However it is interesting to note that the $\pi$ edge mode pairs are a lot more asymmetric
than the $0$ edge mode pairs, where for the latter one does see two peaks, one each
on the left and right edges of the Krylov chain. In contrast the $\pi$ edge mode pairs are strongly peaked at the left edge.
This is because of the highly non-local
structure of the Floquet unitary when a $\pi$ mode exists. Recall that in the previous section
we showed that when a SPM exists, the Floquet unitary is proportional to
the non-local operators ${\mathcal D}$ (SPM phase)  or $\sigma^x_1\sigma^x_L$ (SZM-SPM phase), and therefore has weight on both ends of
the physical chain. Thus
the very first Lanczos step, even when it starts out with an operator localized
at the left edge, does end up ``seeing'' the right end of the chain, resulting in the manifestation of the right edge mode
on the left edge of the Krylov chain.

\subsection{Majorana basis vs spin basis}

Since the Floquet Hamiltonian can always be modified by shifting the quasi-energies by integer multiples of $2\pi$, such
shifts lead to different Krylov Hamiltonians, but ultimately the same physics. In this section we discuss this aspect in detail using the binary drive as
an example. This issue was also discussed in Ref.~\onlinecite{Yates21a}, both for the binary drive and the interacting ternary drive, 
and it was pointed out that a particular choice of branchcut for $\ln(U)$ gives a Krylov Hamiltonian that is
more easy to interpret. This particular choice is one where all the quasi-energies $\epsilon$ lie within the FBZ
defined by $\epsilon T \in \left[-\pi, \pi\right]$.

We now discuss different ways to construct the Krylov Hamiltonian for the binary drive. One is working in the Majorana
basis, and the second is working in the spin basis. Moreover, working in the spin basis, we will discuss two different Krylov
chains, one where the spectrum is unfolded, while the other where the spectrum is folded into the FBZ. Note that in the Majorana basis,
the spectrum is already naturally folded in the FBZ. This is because we have direct access to the orthogonal matrix $K$ whose eigenvalues are pure phases.
Moreover, the eigenvalues of the Krylov Hamiltonian in the Majorana basis equal those of $i\ln{K}$. With a choice of branchcut of
the logarithm along $(-\infty,0)$, this gives a spectrum for the Krylov Hamiltonian that is bounded in the FBZ.

\begin{figure*}
  \includegraphics[width = .99\textwidth]{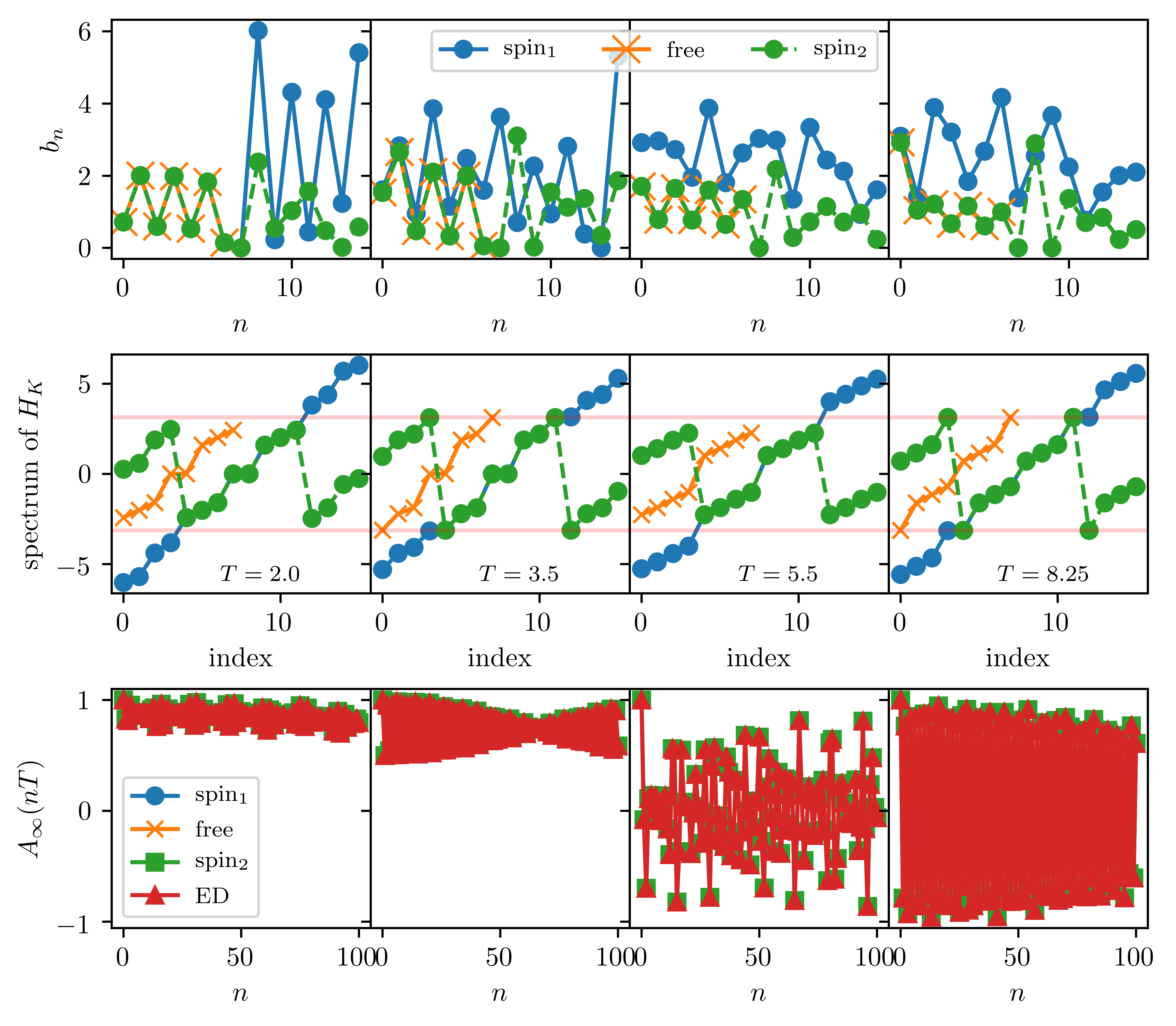}
  \caption{Exploring the difference between the spin
    basis and the Majorana basis for the binary drive with $g=0.3$ and $L=4$.
    All rows, from left to right \(T = 2.0, 3.5,5.5,8.25\).
    These correspond respectively to SZM, SZM-SPM, trivial, and SPM phases. Top rows show the $b_n$s
    for the Majorana basis (labeled as ``free''), and the two spin bases (labeled as spin$_{1,2}$), where
    spin$_1$ is the basis that gives the unfolded spectrum, while spin$_2$ is the basis that gives the spectrum folded into the
    FBZ. 
    Middle rows show the spectra of the corresponding Krylov Hamiltonians, with
    the $x$-axis label ``index'' labeling the eigenvalues.
    The bottom rows plot $A_{\infty}$ obtained from performing
    the time-evolution using the three different Krylov Hamiltonians.
    The red data labeled as ``ED'' in the bottom rows is $A_{\infty}$ from ED.}
  \label{fig4}
\end{figure*}

Fig.~\ref{fig4}
constructs the Krylov Hamiltonian for the binary drive in a few different ways. The
first method is to work in the Majorana basis and to use the
Lanczos method where the generator of stroboscopic dynamics is $i\ln{K}$, with
$K$ given in Eq.~\eqref{eqM}.
For this case, the Lanczos method is a simple basis rotation
and the spectrum of the Krylov Hamiltonian $H_K$, will be exactly the
same as that of $i\ln{K}$. That is, we begin with a single-particle
propagator $i\ln{K}$, and we end with a single-particle propagator $\mathcal{L}$.
The orange data labeled as "free'' in Fig.~\ref{fig4} corresponds to this case where all the computations
are performed in the Majorana basis. The top panels of Fig.~\ref{fig4} shows the hopping parameters obtained from this
procedure, and these are identical to the top panels of Fig.~\ref{fig2}. The middle panels show the corresponding spectra
and are identical to the lower panels of Fig.~\ref{fig2}.
Note that for a system size of $L$, the Hilbert space is exhausted at $2L$ in the Majorana basis. We are working with a system size of $L=4$.
Hence top panels show orange data
that correspond to a total of $2L-1 =7$ hoppings, while middle panels show $2L=8$ eigenvalues.
The lower panels show $A_{\infty}$ as defined in Eq.~\eqref{eqAL} for this Krylov Hamiltonian. They agree perfectly
with the ED calculation (the orange data is masked by the ED data in the lower panels). 

The second way to perform the Lanczos iteration is  to work in the
many-body or spin basis, even though the problem is free. In this case, we are
concerned with the branch of the many-body Floquet Hamiltonian $i\ln{U}$, and not of the single-particle
Hamiltonian $i\ln{K}$.
In Fig.~\ref{fig4}, the blue data set labeled as ``spin$_1$'' corresponds to Lanczos performed
with $T H_F=i\ln{U}$ without any special treatment for the single-particle
status of the problem. The seed state is $\sigma_1^x$.  We see that the resulting spectrum
of $H_K$ in the middle panel
extends beyond the FBZ as indicated by the horizontal red lines which
pass through energies $\pm \pi$. The
$b_n$ are shown in the top row and share little in common with the $b_n$
from the Majorana or free basis (orange crosses). The main differences between the blue data and the orange data is as follows:
the spectrum is unfolded for the former and folded for the latter. Secondly, the former shows 3 gaps (at $\epsilon T =0,\pi,-\pi$), while the latter shows only
two gaps ($\epsilon T=0, \pi$ with $\pi$ continuously connected to $-\pi$).
This difference is due to the fact that the blue data does not have information regarding the periodicity of $U$, and so
the gaps at $\pi$ and $-\pi$ appear to be distinct. 
The orange data has information about the periodicity of the spectrum as it is obtained from the exact $K$. Despite these differences,
both Krylov Hamiltonians generate identical autocorrelation functions, as can be seen in the lower panels of Fig.~\ref{fig4}.

We now wrap the blue spectrum of $H_K$ such that it fits within
the FBZ. This is now given by the green data set labeled ``spin$_2$'' in the middle panels. This folding requires
transforming the Krylov Hamiltonian as follows \cite{Yates21a} $H_K \rightarrow U_K \hat{\epsilon}_{\rm FBZ} U_K^{\dagger}$, where $\hat{\epsilon}_{\rm FBZ}$
is a diagonal matrix where all the energies lie in the FBZ, and $U_K$ is the unitary matrix
that diagonalizes $H_K$ before the folding. 
After this folding, the Krylov Hamiltonian is no longer tri-diagonal, and one
needs to perform a second Lanczos iteration to return to a tri-diagonal Krylov subspace.
The resulting
$b_n$ of this new $H_K$ are shown in the top panel in green, and are labeled as ``spin$_2$''. We note that
this wrapping of the spectrum requires us to fully diagonalize the problem,
which will typically not be possible for us when dealing with large system
sizes. We see that the green $b_n$
of the top row match the $b_n$ obtained from the single-particle Majorana basis upto $b_{n=2L-1}$.
At $n=2L$, $b_{2L}=0$ so that all the $b_n$ appearing after that
represent a second chain decoupled from the part that lies between $n=1\ldots 2L-1$.
The significance of these decoupled chains on the spectrum is shown in the middle panels (green data).
Here we see that the spectrum is repeated twice
for our example.
This repetition does not influence the physics, as can be seen in the lower panel which
plots $A_{\infty}$. This is because once $b_m=0$, the remaining $b_{n>m}$ do not affect the dynamics.
The bottom panel in Fig.~\ref{fig4} plots $A_{\infty}$ for the Krylov Hamiltonian obtained after the folding.
The results agree perfectly with ED, and the two other Krylov Hamiltonians discussed previously.

Thus while the $b_n$ and the spectra (top and center
rows of Fig.~\ref{fig4}) are sensitive to whether the Lanczos is performed on the level
of the many-body or spin basis versus the single-particle or Majorana basis, the dynamics
is not sensitive to it. As touched upon in the previous paragraph, to align the Majorana
picture and the spin picture, we needed to wrap the spectrum of
$H_K$, requiring the full diagonalization of $H_K$. This is only possible
in the free case where we can easily find the full Krylov subspace of
the seed operator. Moreover, the reason for the very different $b_n$s in the spin and in the Majorana bases,
arises due to the ambiguity associated with
the branch of $H_F = i\ln(U)/T$. To avoid this ambiguity, it is better to set up the problem
in an alternate Krylov subspace that involves working directly with the
Floquet unitary $U$, rather than with the Floquet Hamiltonian $H_F$. We discuss this alternate approach in
the next section.

\section{Krylov chain from the Arnoldi iteration} \label{Arnoldi}

We now discuss a different Krylov subspace, one that
arises from the action of $U$ on the seed operator and is known as the Arnoldi iteration \cite{Arnoldi51}.
This differs from the previous section where the generator
of the dynamics was a Hermitian operator, the Floquet Hamiltonian $ T H_F=i\ln{U}$.
Instead, we now work with a unitary operator, and below we outline the steps for obtaining the
corresponding Krylov subspace.

Instead of \(\mathcal{L}\), we now have the quantity $W$
which is defined as follows
\begin{align}
  W |O) &= U^\dagger \hat{O} U\\
  W^n|O) &= \left[U^\dagger\right]^n \hat{O}\left[U\right]^n.
\end{align}
$W$ is unitary because
\begin{align}
  W^\dagger W |O) &= U (U^\dagger \hat{O} U) U^\dagger\\
  &= \hat{O} = W W^\dagger |O).
\end{align}
We now outline the Arnoldi method. This reduces to the Lanczos scheme outlined in
the previous section when $W$ is Hermitian. In particular we will see that the unitarity of
$W$  will no longer produce a simple tri-diagonal matrix. Hermiticity is needed for
the appearance of a tri-diagonal form as we saw in the previous section on the Lanczos iteration
scheme.

Let \(|1) = \sigma_1^x \), then assuming we have found the orthonormal
basis vectors \(|n),|n-1),\dots \), we find \(|n+1)\) by time
evolving it, and projecting out the overlaps with the known basis vectors. Thus we have
\begin{align}
  |n+1') &= W|n) - \sum_{l= 1}^{n}|l)(l|W|n)\nonumber\\
  &= W|n) - \sum_{l = 1}^n w_{l,n} |l)\label{eqW1}\\
  &= \left[ 1 - \sum_{l = 1}^{n}|l)(l| \right] W |n) = P_nW |n).
\end{align}
Above $w_{l,n}= (l|W|n)$ and $P_n=1 - \sum_{l = 1}^{n}|l)(l|$ projects out overlaps with the previously
calculated basis vectors.
Following this we  normalize \(|n+1')\),
\begin{equation}
  |n+1) = \frac{|n+1')}{\sqrt{(n+1'|n+1')}}.\label{eqW2}
\end{equation}
We note that
\begin{align}
  (n+1'|n+1') &= (n+1'|P_n W|n) = (n+1'|W |n),
\end{align}
because $(n+1'|l) =0$ for $l=1\ldots n$ due to the
projector $P_n$. Using Eq.~\eqref{eqW2} the above becomes
\begin{align}
(n+1'|n+1') = \sqrt{(n+1'|n+1')}(n+1|W|n),
\end{align}
implying that
\begin{align}
\sqrt{(n+1'|n+1')} &= (n+1|W|n) = w_{n+1,n}.\label{subd}
\end{align}
Thus we start with $|1)$, find $W|1)$ and $(1|W|1) = w_{1,1}$. Following this
we find the unnormalized $|2') = W|1) - w_{1,1} |1)$.
Then we determine the coefficient $w_{2,1}$ from $(2'|2') = w_{2,1}^2$. This first step fills
in the first two rows of the first column of $W$ in the Arnoldi basis.

Flipping Eq.~\eqref{eqW1} around, we have
\begin{equation}
	W |n) = w_{n+1,n} |n+1) + \sum_{l = 1}^n w_{l,n} |l).
\end{equation}
Thus as we iterate through this Arnoldi algorithm, we will find that
$W$ has the following upper Hessenberg form
\begin{equation}\label{eqWgen}
 W = \begin{pmatrix}
    w_{1,1} & w_{1,2} & \dots \\
    w_{2,1} & w_{2,2} & \dots \\
    0 & w_{3,2} & \dots \\
    0 & 0 & \ddots
  \end{pmatrix}.
\end{equation}
An upper Hessenberg matrix is a square matrix whose elements $w_{i,j}=0$ for $i>j+1$, i.e,
it is a matrix whose elements are zero below the first sub-diagonal \cite{Matrixbook}. In addition,
when the seed operator is Hermitian, all the elements of $W$ are real because under time-evolution
a Hermitian operator stays Hermitian, with the matrix elements of $W$ simply denoting the weight
of different Hermitian operators at a particular step in the iterative procedure. Typically the structure of
$W$ is such that the elements on the first sub-diagonal are dominant. These elements besides measuring
the norm of the operators (see Eq.~\eqref{subd}), also measure how much the operator is spreading into
new parts of the Krylov subspace. In contrast, all other elements of $W$ are simply the overlap of the new operator
generated at each iteration, with previous elements
of the Krylov subspace.
Moreover, the further an element is from the
first sub-diagonal, the smaller it is. This observation will be handy later when we derive analytic expressions
for $W$.

The autocorrelation function in terms of $W$ now has the form
\begin{equation}\label{ainfWN}
  \ainf(nT) = (1|W^n|1).
\end{equation}
Just as for $\mathcal{L}$, the iteration must be cut short after $N$
steps. Due to the truncation, the resulting $W$
will not be unitary.
As we will show later in the context of ASZMs and ASPMs,
as long as $N$ is sufficiently large, the truncated $W$ does well in reproducing the dynamics.
The success of the approximation is a good indication
that if the edge modes are sufficiently localized at the edge of the Krylov chain, the truncation scheme
does not affect the physics.

We note that for the last state $|N)$ we have,
\begin{equation}
  W |N) = \sum_{l =1}^N w_{l,N} |l) + w_{N+1,N}|N+1),
\end{equation}
and by stopping at $|N)$, we are making some form of a Markov
approximation at a rate of $w_{N+1,N}$. Thus we expect almost all
operator dynamics produced by a truncated $W$ to eventually cause a decay in time.

\subsection{Strong modes in limiting cases}
We now discuss the form of the $W$ matrix for some analytically tractable
limiting cases.  These limits were already discussed in previous sections in the context of the matrix $K$ in Eq.~\eqref{eqM}.
\begin{itemize}
\item{\(Tg = (2n+1)\pi\), \(T\) arbitrary:} For this case \(e^{-iTgH_z/2}
  \propto \mathcal{D}\), thus \(U^\dagger \sx U =  \mathcal{D}\sx \mathcal{D} = -\sx,\) and a SPM exists, whereas
  there is no SZM. The first step of the Arnoldi iteration is \(|1) =
  \sx\) and the next state is
  \begin{equation}
    |2') = W |1) = -|1).
  \end{equation}
  Thus the Krylov subspace terminates after the first basis vector
  $|1) = \sx $, and \(W\) is a "matrix" with a single element,
  \(-1\).

\item{\(Tg = (2n)\pi\), \(T\) arbitrary:} For this case
  \(e^{-iTgH_z/2}
  \propto 1\) and \(U^\dagger \sx U = \sx\). We have a SZM, but no SPM.
  Following the same steps as in the previous case, it is clear
  that \(W = 1\).

\item{\(T = (2n+1)\pi\), \(Tg\) arbitrary:} We now have \(e^{- i
  TH_{xx}/2} \propto \sigma_1^x \sigma_L^x\), with \(e^{-i T
  gH_z/2} = \prod_l \left[ \cos (T g/2) - i \sigma_l^z \sin(T
  g/2)\right]\). For this case
  Eq.~\eqref{eqU3} holds. We now construct the Krylov subspace for \(\sx\).
  Beginning with
  \(|1) = \sx\),  we have using Eq.~\eqref{eqU3}
  \begin{equation}
    W |1) = \cos (Tg) \sx  - \sin(Tg) \sy,
  \end{equation}
  giving,
  \begin{equation}
    |2') = W |1) - \cos(Tg) |1) = - \sin(Tg) \sy.
  \end{equation}
After normalization we obtain
  \begin{equation}
    |2) = -\sy.
  \end{equation}
  Moreover, again using Eq.~\eqref{eqU3}, we can see that
  $W |2)$ will be a linear combination of $|1), |2)$, and in particular
  \begin{equation}
    W =
    \begin{pmatrix}
      \cos (Tg) & |\sin(Tg)| \\
      |\sin(Tg)| & -\cos(Tg)
    \end{pmatrix},
  \end{equation}
which corresponds to having both a SZM and a SPM because the eigenvalues of $W$ are $\pm 1$.

\item{\(T= 2n\pi\) and \(Tg \) arbitrary:} For this case we have \(e^{-i
  TJ_xH_{xx}/2} \propto 1\) and the time-evolution is given by Eq.~\eqref{eqU4}.
  Carrying out the same steps as before one finds that
  the Krylov subspace for \(\sx\) is two
  dimensional with
\begin{equation}
    W =
    \begin{pmatrix}
      \cos (Tg) & -|\sin(Tg)| \\
      |\sin(Tg)| & \cos(Tg)
    \end{pmatrix}.
  \end{equation}
Thus the eigenvalues are pure phases that do not equal $\pm 1$, unless $Tg = n \pi $, where $n$ is an integer.
When $n\in $ even, we have a SZM, and when $n \in$ odd, we have a SPM.
\end{itemize}

\section{Arnoldi iteration for the binary drive} \label{Arnoldi-free}

In this section we present the analytic form of $W$ for the binary drive, and
discuss it for some limiting cases which are more general than those
discussed in the previous section. This discussion is aimed at showing
the connection between strong modes and the topological properties
of the Hamiltonian $i\ln{W}$.

Recall that for the binary drive,
we can work in the Majorana basis where a chain of length $L$ requires diagonalizing the problem
in a reduced Hilbert space of size $2L$.
It is straightforward, but tedious to verify that for a chain of length $L=4$
the Arnoldi alogrithm produces the following
$8\times 8$ $W$ matrix,
\begin{widetext}
  \begin{equation}\label{eqW}
  W=  \begin{pmatrix}
      c_1 & - s_1 c_2 & s_2 s_1 c_1 & - s_2 s_1^{2} c_2 & s_2^{2}
      s_1^2 c_1 & - s_2^2 s_1^3 c_2 & s_2^3 s_1^3 c_1 & - s_2^3
      s_1^4\\ s_1 & c_2 c_1 & - s_2 c_1^2 & s_1c_1s_2 c_2 &-s_2^2 s_1
      c_1^2 & s_2^2 s_1^2 c_2 c_1 & - s_2^3 s_1^2 c_1^2 & s_2^3 s_1^3
      c_1\\ 0 & s_2 & c_2 c_1 & - s_1 c_2^2 & s_1c_1s_2 c_2 & - s_2
      s_1^2 c_2^2 & s_2^2 s_1^2 c_2 c_1 & - s_2^2 s_1^3 c_2\\ 0 & 0 &
      s_1 & c_2 c_1 & - s_2 c_1^2 & s_1c_1s_2 c_2 & - s_2^2 s_1 c_1^2
      & s_2^2 s_1^2 c_1\\ 0 & 0 & 0 & s_2 & c_2 c_1 & - s_1 c_2^2 &
      s_1c_1s_2 c_2 & - s_2 s_1^2 c_2\\ 0 & 0 & 0 & 0 & s_1 & c_2 c_1
      & - s_2 c_1^2 & s_1c_1s_2 \\ 0 & 0 & 0 & 0 & 0 & s_2 & c_2 c_1 &
      - s_1 c_2\\ 0 & 0 & 0 & 0 & 0 & 0 & s_1 & c_1
    \end{pmatrix},
  \end{equation}
where $c_{1,2}, s_{1,2}$ were defined in Eq.~\eqref{eqcsdef}. We discuss the structure of $W$
as it will help us make further approximations.  Firstly since $W$ is unitary, its rows and columns form
an orthonormal basis. Secondly,
for even rows, starting on the column corresponding to the
sub-diagonal, we have
\begin{equation}
    \{s_1,\ c_1 c_2,\ -s_2 c_1^2,\ (s_1 s_2) c_1 c_2,\ - (s_1 s_2) s_2 c_1^2,\
    \dots,\ (s_1 s_2)^{n} (c_1 c_2),\ (s_1 s_2)^n (-s_2 c_1^2),\ \dots \}.
  \end{equation}
Thirdly, for odd rows, starting on the column corresponding to the
sub-diagonal, we have
  \begin{equation}
    \{s_2,\ c_1 c_2,\ -s_1 c_2^2,\ (s_1 s_2) c_1 c_2,\ - (s_1 s_2) s_1 c_2^2,\
    \dots,\ (s_1 s_2)^{n} (c_1 c_2),\ (s_1 s_2)^n (-s_1 c_2^2),\ \dots \}.
  \end{equation}
  The rows and columns at the edge of the matrix can be determined from
  using the above rules and then dividing by $c_2$. This also ensures orthonormality of the rows
  and columns. In what follows we will further explore $W$, and in particular $\ln{W}$ for
  limiting cases that give SZMs and SPMs. The reason for exploring $\ln{W}$ is that
  it has a Hamiltonian description, and therefore the topological properties can be
  more easily discerned.

\subsection{$\ln{W}$ for $g T \ll 1$, $T\ll 1$: SZM phase}

First let us consider the case when $gT\ll 1$. For this case we have $s_{2},c_{2}$ arbitrary, and $s_1\ll 1, c_1 \approx 1$.
Truncating the $W$ matrix in Eq.~\eqref{eqW} to terms of $O(s_1)$ we have
  \begin{align}
&W=    \begin{pmatrix}
      1 & - s_1 c_2 & s_2 s_1 & 0 & 0 & 0 & 0 & 0\\ s_1 & c_2 & - s_2 & s_1s_2 c_2 &-s_2^2 s_1
       & 0 & 0 & 0\\ 0 & s_2 & c_2 & - s_1 c_2^2 & s_1s_2 c_2 &0 & 0 & 0\\ 0 & 0 &
      s_1 & c_2 & - s_2 & s_1s_2 c_2 & - s_2^2 s_1
      & 0\\ 0 & 0 & 0 & s_2 & c_2  & - s_1 c_2^2 &
      s_1s_2 c_2 & 0\\ 0 & 0 & 0 & 0 & s_1 & c_2
      & - s_2 & s_1s_2 \\ 0 & 0 & 0 & 0 & 0 & s_2 & c_2 &
      - s_1 c_2\\ 0 & 0 & 0 & 0 & 0 & 0 & s_1 & 1
\end{pmatrix} + O(s_1^2) = W_0 + \delta W,\\
& W_0= \begin{pmatrix}
      1 & 0 & 0 & 0 & 0 & 0 & 0 & 0\\ 0 & c_2 & - s_2 & 0 &0
       & 0 & 0 & 0\\ 0 & s_2 & c_2 & 0& 0 &0 & 0 & 0\\ 0 & 0 &
      0 & c_2 & - s_2 & 0 &0
      & 0\\ 0 & 0 & 0 & s_2 & c_2  & 0 &
      0 & 0\\ 0 & 0 & 0 & 0 & 0 & c_2
      & - s_2 & 0 \\ 0 & 0 & 0 & 0 & 0 & s_2 & c_2 &
      0\\ 0 & 0 & 0 & 0 & 0 & 0 & 0 & 1
\end{pmatrix}= W(g T=0),\\
&\delta W=  \begin{pmatrix}
      0 & - s_1 c_2 & s_2 s_1 & 0 & 0 & 0 & 0 & 0\\ s_1 & 0 & 0 & s_1s_2 c_2 &-s_2^2 s_1
       & 0 & 0 & 0\\ 0 & 0 & 0 & - s_1 c_2^2 & s_1s_2 c_2 &0 & 0 & 0\\ 0 & 0 &
      s_1 & 0 & 0 & s_1s_2 c_2 & - s_2^2 s_1
      & 0\\ 0 & 0 & 0 & 0 & 0  & - s_1 c_2^2 &
      s_1s_2 c_2 & 0\\ 0 & 0 & 0 & 0 & s_1 &0
      & 0 & s_1s_2 \\ 0 & 0 & 0 & 0 & 0 & 0 & 0 &
      - s_1 c_2\\ 0 & 0 & 0 & 0 & 0 & 0 & s_1 & 0
\end{pmatrix}.
\end{align}
  Above $W_0 = W(g T=0)$, while $\delta W$ is the leading correction about this
  point and is $O(s_1)$. If in addition we impose $T\ll 1$ which is equivalent to $s_2\ll 1$, the matrices
$W_0$ and $\delta W$ commute if terms of $O(s_1s_2), O(s_1^2), O(s_2^2)$ and higher are dropped.
Recall that if $A$ and $B$ are two commuting matrices $\left[A,B\right]=0$, then,
\begin{align}
  \ln{(A+B)} \approx \ln\left[A\right] + \left[A\right]^{-1} B + \ldots. \label{eqlogexp}
\end{align}
Using,
Eq.~\eqref{eqlogexp} and setting $A = W_0, B = \delta W$, 
we find the following expressing to first order in $s_1$ and $T$
\begin{align}
i\ln\left[W\right] \approx     \left(
\begin{array}{cccccccc}
 0 & -is_1 & 0 & 0 & 0 & 0 & 0 & 0 \\
 is_1 & 0 & -iT & 0 & 0 & 0 & 0 & 0 \\
 0 & iT & 0 & -is_1 & 0 & 0 & 0 & 0 \\
 0 & 0 & is_1 & 0 & -iT & 0 & 0 & 0 \\
 0 & 0 & 0 & iT & 0 & -is_1 & 0 & 0 \\
 0 & 0 & 0 & 0 & is_1 & 0 & -iT & 0 \\
 0 & 0 & 0 & 0 & 0 & iT & 0 & -is_1 \\
 0 & 0 & 0 & 0 & 0 & 0 & is_1 & 0 \\
\end{array}
\right) .   \label{eqlnWzm}
\end{align}
Above $i\ln{W}$ represents a Floquet Hamiltonian that is
effectively a SSH model with topologically non-trivial sign of the dimerization for $|s_1|<T$. Thus a zero mode
is guaranteed. This shows that the SZM of the original Hamiltonian manifests as an edge mode of
a topologically non-trivial Hamiltonian $i\ln{W}$ in the single particle Krylov
subspace.

\subsection{$\ln{W}$ for $g T \approx \pi$, $T\ll 1$: SPM phase}

Let us first consider $g T \approx \pi$ where $s_1\ll 1$ and $c_1=-1$. Truncating $W$ in Eq.~\eqref{eqW} to $O(s_1)$ we obtain
  \begin{align}
&W=    \begin{pmatrix}
      -1 & - s_1 c_2 & -s_2 s_1 & 0 & 0 & 0 & 0 & 0\\ s_1 & -c_2 & - s_2 & -s_1s_2 c_2 &-s_2^2 s_1
       & 0 & 0 & 0\\ 0 & s_2 & -c_2 & - s_1 c_2^2 & -s_1s_2 c_2 &0 & 0 & 0\\ 0 & 0 &
      s_1 & -c_2 & - s_2 & -s_1s_2 c_2 & - s_2^2 s_1
      & 0\\ 0 & 0 & 0 & s_2 & -c_2  & - s_1 c_2^2 &
      -s_1s_2 c_2 & 0\\ 0 & 0 & 0 & 0 & s_1 & -c_2
      & - s_2 & -s_1s_2 \\ 0 & 0 & 0 & 0 & 0 & s_2 & -c_2 &
      - s_1 c_2\\ 0 & 0 & 0 & 0 & 0 & 0 & s_1 & -1
\end{pmatrix} + O(s_1^2)= W_0+ \delta W,\\
& W_0= W(gT=\pi)= -\begin{pmatrix}
      1 & 0 & 0 & 0 & 0 & 0 & 0 & 0\\ 0 & c_2 & s_2 & 0 &0
       & 0 & 0 & 0\\ 0 & -s_2 & c_2 & 0& 0 &0 & 0 & 0\\ 0 & 0 &
      0 & c_2 & s_2 & 0 &0
      & 0\\ 0 & 0 & 0 & -s_2 & c_2  & 0 &
      0 & 0\\ 0 & 0 & 0 & 0 & 0 & c_2
      & s_2 & 0 \\ 0 & 0 & 0 & 0 & 0 & -s_2 & c_2 &
      0\\ 0 & 0 & 0 & 0 & 0 & 0 & 0 & 1
\end{pmatrix},\nonumber\\
&\delta W=  \begin{pmatrix}
      0 & - s_1 c_2 & -s_2 s_1 & 0 & 0 & 0 & 0 & 0\\ s_1 & 0 & 0 & -s_1s_2 c_2 &-s_2^2 s_1
       & 0 & 0 & 0\\ 0 & 0 & 0 & - s_1 c_2^2 & -s_1s_2 c_2 &0 & 0 & 0\\ 0 & 0 &
      s_1 & 0 & 0 & -s_1s_2 c_2 & - s_2^2 s_1
      & 0\\ 0 & 0 & 0 & 0 & 0  & - s_1 c_2^2 &
      -s_1s_2 c_2 & 0\\ 0 & 0 & 0 & 0 & s_1 &0
      & 0 & -s_1s_2 \\ 0 & 0 & 0 & 0 & 0 & 0 & 0 &
      - s_1 c_2\\ 0 & 0 & 0 & 0 & 0 & 0 & s_1 & 0
\end{pmatrix}.
  \end{align}
  Above $\delta W$ is $O(s_1)$, and represents the leading correction to $W$ around the point $g T=\pi$.
  If we now also assume $T\ll 1$ which is equivalent to $s_2\ll 1$, then $W_0$ and $\delta W$ commute
  if terms of $O(s_1s_2), O(s_1^2), O(s_2^2)$ and higher are dropped.
Performing the expansion in Eq.~\eqref{eqlogexp} we find, to first order in $s_1$ and $T$
\begin{align}
i\ln\left[W\right] \approx \pm \pi  +  \left(
\begin{array}{cccccccc}
 0 & is_1 & 0 & 0 & 0 & 0 & 0 & 0 \\
 -is_1 & 0 & iT & 0 & 0 & 0 & 0 & 0 \\
 0 & -iT & 0 & is_1  & 0 & 0 & 0 & 0 \\
 0 & 0 & -is_1  & 0 & iT & 0 & 0 & 0 \\
 0 & 0 & 0 & -iT & 0 & is_1  & 0 & 0 \\
 0 & 0 & 0 & 0 & -is_1  & 0 & iT & 0 \\
 0 & 0 & 0 & 0 & 0 & -iT & 0 & is_1 \\
 0 & 0 & 0 & 0 & 0 & 0 & -is_1 & 0 \\
\end{array}
\right).   \label{lnWpi}
\end{align}
Thus we find that the Floquet Hamiltonian $i\ln{W}$
now represents a 
topologically non-trivial SSH model (for $|s_1|<T$), with a constant shift of $\pi$, the latter ensuring that the zero mode
of the SSH model is shifted in energy by $\pi$. A similar picture was discussed in Ref.~\onlinecite{Yates21a}, where the
discussion was presented for the Floquet Hamiltonian $i \ln{K}$.

\subsection{ $\ln{W}$ for $T \approx \pi$: SZM-SPM phase}
Let us consider the case $T \approx \pi$.
For this case we have $c_2\approx -1,s_2\ll 1$ while $c_{1},s_1$ are arbitrary. Truncating Eq.~\eqref{eqW}
to $O(s_2)$ we obtain
  \begin{align}
W&=  \begin{pmatrix}
      c_1 & s_1 & s_1 s_2 c_1 & s_1^2 s_2& 0 & 0 & 0 & 0
   \\ s_1 & -c_1 & -s_2c_1^2 & -s_1s_2 c_1 &0& 0 & 0 & 0
   \\ 0 & s_2 & -c_1 & - s_1 & -s_1s_2 c_1 &-s_1^2s_2 & 0 & 0
   \\ 0 & 0 & s_1 & -c_1 & - s_2c_1^2 & -s_1s_2 c_1 &0 & 0
   \\ 0 & 0 & 0 & s_2 & -c_1  & - s_1 &-s_1s_2 c_1 & s_1^2s_2
   \\ 0 & 0 & 0 & 0 & s_1 & -c_1 & - s_2c_1^2 & s_1s_2c_1
   \\ 0 & 0 & 0 & 0 & 0 & s_2 & -c_1 &s_1
   \\ 0 & 0 & 0 & 0 & 0 & 0 & s_1 & c_1
\end{pmatrix}+ O(s_2^2) = W_0 + \delta W,\\
 W_0&=W(T=\pi)=\begin{pmatrix}
      c_1 & s_1 & 0 & 0& 0 & 0 & 0 & 0
   \\ s_1 & -c_1 & 0 & 0 &0& 0 & 0 & 0
   \\ 0 & 0 & -c_1 & - s_1 & 0 &0 & 0 & 0
   \\ 0 & 0 & s_1 & -c_1 & 0 & 0 &0 & 0
   \\ 0 & 0 & 0 & 0 & -c_1  & - s_1 &0 & 0
   \\ 0 & 0 & 0 & 0 & s_1 & -c_1 & 0 & 0
   \\ 0 & 0 & 0 & 0 & 0 & 0 & -c_1 &s_1
   \\ 0 & 0 & 0 & 0 & 0 & 0 & s_1 & c_1
    \end{pmatrix}\nonumber\\
    &= \begin{pmatrix}
      & 0 & 0& 0 & 0 & 0 & 0
   \\ \pm i e^{\mp i\frac{\pi}{2}(c_1\sigma_z + s_1\sigma_x)} & 0 & 0 &0& 0 & 0 & 0
   \\ 0 & 0 &  & 0 &0 & 0 & 0
   \\ 0 & 0 & -e^{i T \sigma_y}&0 & 0 &0 & 0
   \\ 0 & 0 & 0 & 0 & &0 & 0
   \\ 0 & 0 & 0 & 0 & -e^{i T \sigma_y}& 0 & 0
   \\ 0 & 0 & 0 & 0 & 0 & 0 &
   \\ 0 & 0 & 0 & 0 & 0 & 0 & \pm i e^{\mp i\frac{\pi}{2}(-c_1\sigma_z + s_1\sigma_x)}
    \end{pmatrix},
    \nonumber\\
\delta W&= \begin{pmatrix}
      0 & 0 & s_1 s_2 c_1 & s_1^2 s_2& 0 & 0 & 0 & 0
   \\ 0 & 0 & -s_2c_1^2 & -s_1s_2 c_1 &0& 0 & 0 & 0
   \\ 0 & s_2 & 0 & 0 & -s_1s_2 c_1 &-s_1^2s_2 & 0 & 0
   \\ 0 & 0 & 0 & 0 & - s_2c_1^2 & -s_1s_2 c_1 &0 & 0
   \\ 0 & 0 & 0 & s_2 & 0  & 0 &-s_1s_2 c_1 & s_1^2s_2
   \\ 0 & 0 & 0 & 0 & 0 & 0 & - s_2c_1^2 & s_1s_2c_1
   \\ 0 & 0 & 0 & 0 & 0 & s_2 & 0 &0
   \\ 0 & 0 & 0 & 0 & 0 & 0 & 0 & 0
\end{pmatrix}.
  \end{align}
  Above $\delta W$ is the leading correction to $W_0$ and is $O(s_2)$.
  
At this stage we simply discuss the Floquet Hamiltonian $i\ln(W_0)$,
\begin{align}
&    \ln\left[W_0\right] \approx \begin{pmatrix}
      & 0 & 0& 0 & 0 & 0 & 0
   \\ \ln\biggl[\pm i e^{\mp i\frac{\pi}{2}(c_1\sigma_z + s_1\sigma_x)}\biggr] & 0 & 0 &0& 0 & 0 & 0
   \\ 0 & 0 &  & 0 &0 & 0 & 0
   \\ 0 & 0 & \ln\biggl[-e^{i T \sigma_y}\biggr]&0 & 0 &0 & 0
   \\ 0 & 0 & 0 & 0 & &0 & 0
   \\ 0 & 0 & 0 & 0 & \ln\biggl[-e^{i T \sigma_y}\biggr]& 0 & 0
   \\ 0 & 0 & 0 & 0 & 0 & 0 &
   \\ 0 & 0 & 0 & 0 & 0 & 0 & \ln\biggl[\pm i e^{\mp i\frac{\pi}{2}(-c_1\sigma_z + s_1\sigma_x)}\biggr]
  \end{pmatrix}.
\label{lnW0pi}
\end{align}
Thus we see that $i\ln{(W_0)}$, for these parameters, has a complex edge structure represented by
the $2\times 2$ block in the upper and lower diagonals. The eigenvalues of these edge modes are $0, \pm \pi$.
Moreover, these edge modes are completely decoupled from a bulk which
has a dimerization of a topologically non-trivial SSH model represented by the repeated sub-block $\ln\biggl[-e^{i T \sigma_y}\biggr]$.
Deviating from this exactly solvable limit will couple the edge modes
weakly to the bulk states, and will also generate longer range hopping. However, these edge modes are protected as long
as the bulk gap remains non-zero. Thus we see that the SZM-SPM phase also can be interpreted as topologically protected edge modes
of a single-particle Hamiltonian in Krylov subspace. The advantage of the discussion in this section is that, even when interactions are non-zero,
one may construct an effectively single-particle $W$. Thus,
topological features, or lack thereof, of $W$ can shed light on the lifetime of the almost strong modes. 

\end{widetext}

\section{Lifetime of edge modes from the Arnoldi Algorithm}\label{LFAA}

In this section we construct the approximate edge mode and derive its lifetime for the most general $W$ matrix whose upper Hessenberg form is
highlighted in Eq.~\eqref{eqWgen}. The results derived here hold for the free and the interacting cases.
Recall that for the free system we have a SZM and/or a SPM. These acquire a lifetime due to finite size effects, i.e, their lifetime
scales exponentially in the system size $L$. On the other hand, when the interactions are non-zero,
the edge modes acquire a system-size independent lifetime provided the system size is large enough. For small system sizes,
there is not much of a difference between strong modes and almost strong modes as both have lifetimes that grow exponentially
with system size.

We denote edge modes of $W$, both exact and approximate,
as $|\psi_{0,\pi}\rangle$, where the subscript indicates whether it is
a zero mode or a $\pi$-mode. We then argue that the lifetime $\gamma_{0,\pi}^{-1}$
of the edge mode is well approximated by the expression
\begin{align}\label{eqlt1}
e^{-\gamma_{0,\pi}} = |\langle \psi_{0,\pi}|W|\psi_{0,\pi}\rangle|,
\end{align}
where $\gamma_{0,\pi}$ is in dimensionless units. We will present numerical evidence for
this.
It is straightforward to see that when $|\psi_{0,\pi}\rangle$ are exact edge modes, $|\langle \psi_{0,\pi}|W|\psi_{0,\pi}\rangle|=1$
and the decay rate $\gamma_{0,\pi}=0$. We will apply the formula to the case where we have quasi-stable edge modes, so that
$|\langle \psi_{0,\pi}|W|\psi_{0,\pi}\rangle|<1$, resulting in a non-zero decay rate.

We solve for the edge modes by determining
the left eigenvectors of $W$,
\begin{align}
\langle \psi_{0,\pi} |W = \tilde{\lambda} \langle \psi_{0,\pi} |,
\end{align}
with $\tilde{\lambda}=1$ for $\langle \psi_0|$ and $\tilde{\lambda}=-1$ for $\langle \psi_{\pi}|$.
Essentially this allows us
to work from the first site and iterate into the bulk.
We denote the elements of $\langle \psi_{0,\pi}|$ by $\psi_i$ where the subscript $i$ is the site index.
The first step involves the inner product of the left eigenvector with
the first column of $W-\tilde{\lambda}$ which gives
\begin{align}
  \psi_{1} &= 1,\\
  \psi_{2} &= -\biggl(\frac{w_{1,1} - \tilde{\lambda}}{w_{2,1}}\biggr).
\end{align}
Above we set $\psi_1=1$ as we are interested in an edge mode.
The inner product of the left eigenvector onto the
second column of the matrix $W - \tilde{\lambda}$ gives
\begin{align}
\psi_{3} &= -\frac{1}{w_{3,2}}
  \left[(w_{2,2} -\tilde{\lambda})\psi_{2} + w_{1,2} \psi_{1} \right].
\end{align}
The pattern is clear for the rest of the eigenvector components,
\begin{equation}
  \psi_{n+1} = -\frac{1}{w_{n+1,n}} \left[ (w_{n,n} - \tilde{\lambda})\psi_{n}
  + \sum_{i = 1}^{n-1} w_{i,n} \psi_i \right],
  \label{eq:recurrance1}
\end{equation}
with a final normalization step at the end of the calculation.

We now rewrite Eq.~\eqref{eq:recurrance1} as follows as it will be helpful
later
\begin{equation}\label{eqrec2}
  \sum_{i =  1}^{n-1} \psi_{i} w_{i,n}  = -\psi_{n+1} w_{n+1,n} -(w_{n,n} - \tilde{\lambda})
  \psi_{n}.
\end{equation}
Since all the elements of $W$ are real, the components  $\psi_i$ of the edge mode are also real.
We now set out to compute the right hand side of Eq.~\eqref{eqlt1}. It is staightforward to see that
\begin{widetext}
\begin{align}
  \langle \psi_{0,\pi} |W | \psi_{0,\pi} \rangle &=
  \sum_{k = 1}^{N-1} \left[
    \psi_{k+1} w_{k+1,k} + \psi_{k} w_{k,k}+
    \sum_{m = 1}^{k-1} \psi_m w_{m,k}
    \right] \psi_k
  + \sum_{m = 1}^N \psi_m w_{m,N} \psi_N\\
  &=
  \sum_{k = 1}^{N-1} \left[
    \psi_{k+1} w_{k+1,k} + \psi_{k} w_{k,k}+
    \left(-\psi_{k+1}w_{k+1,k} - (w_{k,k} - \tilde{\lambda})\psi_k \right)
    \right] \psi_k
  + \sum_{m = 1}^N \psi_m w_{m,N} \psi_N\\
  &=
  \sum_{k = 1}^{N-1}\tilde{\lambda}\psi_k^2
  + \sum_{m = 1}^N \psi_m w_{m,N} \psi_N\\
  &=\tilde{\lambda} \left[
  \sum_{k = 1}^{N}\psi_k^2
  +\tilde{\lambda} \sum_{m = 1}^N \psi_m w_{m,N} \psi_N - \psi_N^2\right].
\end{align}
\end{widetext}
In the second equality above we have used Eq.~\eqref{eqrec2} to simplify the expression.

We normalize $\psi_{0,\pi} \rightarrow \psi_{0,\pi}/ N_{0,\pi}$ and use that $|\tilde{\lambda}|=1$ to obtain
\begin{equation}
  | \langle \psi_{0,\pi} | W |\psi_{0,\pi} \rangle |=
  1 + \frac{\tilde{\lambda}}{N_{0,\pi}^2} \sum_{m = 1}^N \psi_m w_{m,N} \psi_N
  -\frac{\psi_N^2}{N_{0,\pi}^2}.
\end{equation}
Using Eq.~\eqref{eqlt1}, we identify the decay rate from Taylor expanding the left hand side of
the same equation to obtain
\begin{equation}
  \gamma_{0,\pi} \approx \frac{\psi_N^2}{N_{0,\pi}^2} - \frac{\tilde{\lambda}}{N_{0,\pi}^2}
  \sum_{m = 1}^N \psi_m w_{m,N} \psi_N.
  \label{eq:lifetime2}
\end{equation}
We cannot proceed further without making approximations to $W$ or to
$\psi$. As it stands, Eq.~\eqref{eq:lifetime2} agrees well with
the quantity
$\ln{|\langle \psi_{0,\pi}|W|\psi_{0,\pi}\rangle|}$ which was how the decay-rate was defined in Eq.~\eqref{eqlt1}.
We give evidence of this in the next section.
Empirically, we also find that the second term in Eq.~\eqref{eq:lifetime2} is often much smaller than
the first, thus we can approximate the expression for the decay rate
even further as
\begin{equation}\label{eqlifetime3}
  \gamma_{0,\pi} \approx \psi_N^2.
\end{equation}
Above we are explicitly working with the normalized wavefunction and therefore we have dropped the factor of $N_{0,\pi}^2$.

Thus the decay rate is the probability of the edge mode to be located at the site $N$. Recall that
$N$ is the dimension of the $W$ matrix. For the free case, if $N=2L$, this exhausts the Hilbert space.
For the interacting problem $N$ has to be very large $\approx e^{2L}$ in order to exhaust the Hilbert space.
Thus when working with interacting systems, we will always be truncating $W$. However as long as this truncation occurs
for an $N$ which is large enough, and the quasi-stable edge mode is sufficiently localized at the edge, the decay rate $\gamma_{0,\pi}$,
and therefore the probability of finding the particle at
site $N$, will be independent of $N$ (for a given system size $L$).

\section{Krylov subspace with Interactions} \label{Int}

We now consider the interacting example by modifying the binary drive in Eq.~\eqref{eqU1} into a ternary drive where
the third part of the drive breaks the free fermion nature of the problem. Specifically we study
\begin{equation}
  U = e^{-i \frac{T}{2} J_z H_{zz}} e^{-i \frac{T}{2} J_x H_{xx}} e^{-i \frac{T}{2} g H_z}, \label{eqU2}
\end{equation}
where the interacting part is
\begin{align}
  H_{zz} = \sum_{i=1}^{L-1} \sigma^z_i \sigma^z_{i+1}.
\end{align}
The above drive was studied in detail in \cite{Yates19,Yates21a}, and it was found that the SZMs and SPMs
get modified to ASZMs and ASPMs where the latter are characterized by a system size independent
lifetime, for sufficiently large system sizes. Moreover, this lifetime is still  very long as compared
to the time needed for the bulk of the system to heat to infinite temperature. These studies were carried out
employing the Lanczos iteration scheme. Therefore, we discuss the role of interactions only within the Arnoldi iteration scheme.

\subsection{Krylov subspace from the Arnoldi iteration}

Section \ref{Arnoldi} outlined the Arnoldi procedure. The key quantity is the unitary matrix $W$ in Eq.~\eqref{eqWgen}
which has a characteristic upper Hessenberg form. The analytic expression for $W$ and $\ln{W}$
for the binary drive was presented in Section \ref{Arnoldi-free}, and the
lifetime of the edge mode operators for the general case including interactions
was derived in Section \ref{LFAA}. In this section we present the $W$ matrix for the
interacting problem, and apply the results for the lifetime, derived in Section \ref{LFAA}.

\begin{figure}
  \includegraphics[width = .49\textwidth]{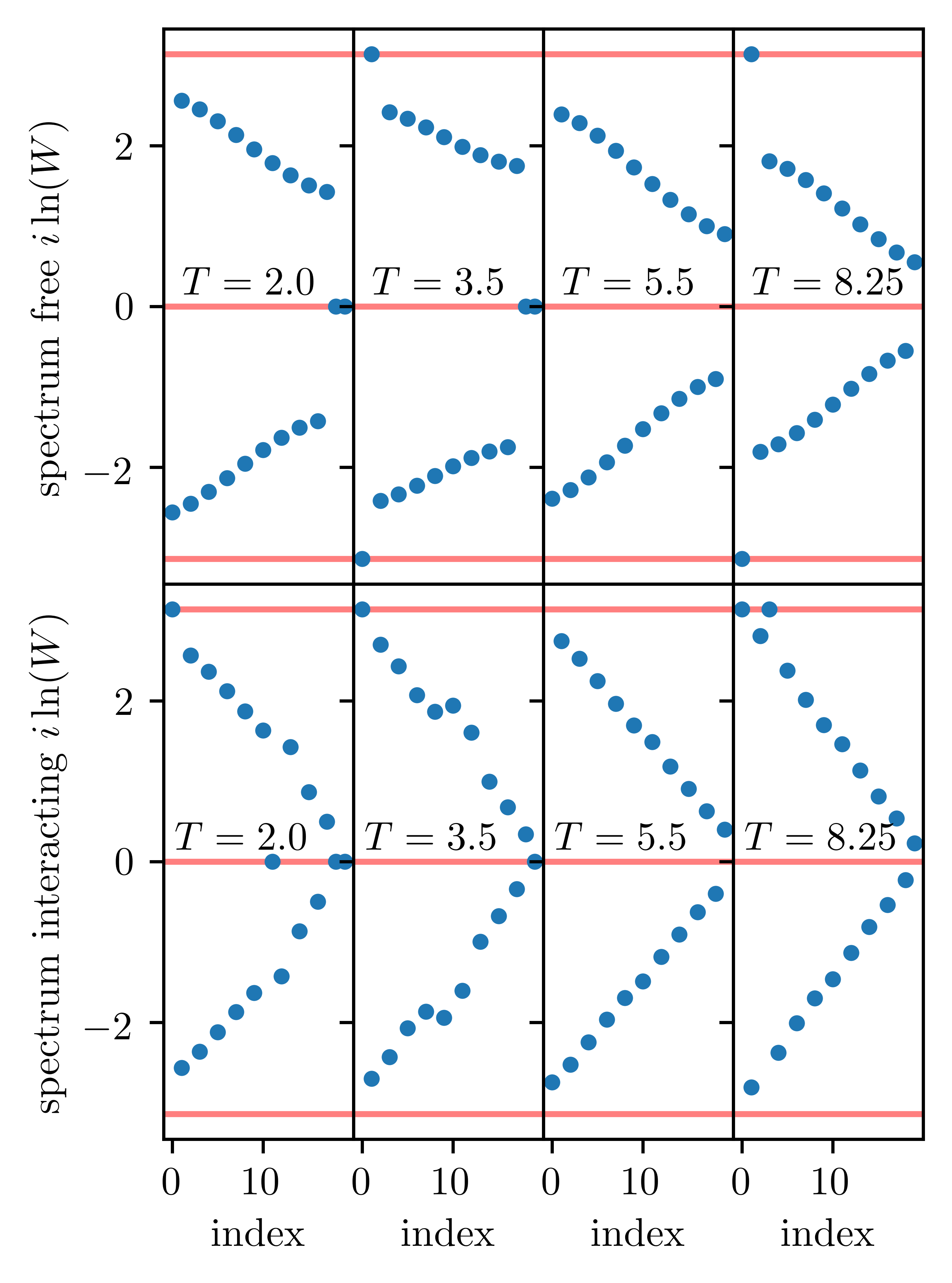}
  \caption{Spectrum of  $i\ln{W}$ for  free example (upper row) and an
    interacting example with $J_z=0.05$ (lower row).  $x$-axis label ``index'' denotes the location of the eigenvalues arranged in descending order
    of their magnitude.
    For all data $L=10$ and $W$
    is a $20 \times 20$ matrix. In addition $g=0.3$. The four different values of $T$ are
    from left to right \(T = 2.0, 3.5,5.5,8.25\).
    These correspond respectively to (A)SZM, (A)SZM-(A)SPM, trivial, and (A)SPM phases. The horizonal
  red lines indicate $0,\pm \pi$.}
  \label{fig7}
\end{figure}

Fig.~\ref{fig7} shows the spectrum of $i\ln{W}$ for the free case (top panel) and for the interacting case (bottom panel)
where $J_z=0.05$ for the interacting case.
All other parameters are common between the top and bottom panels. In particular, $g=0.3$, the system size is $L=10$, and
$W$ is a $20\times 20$ matrix. Thus for the free case, $W$ is exact. However for the interacting case, $W$ is not exact because
the Hilbert space is larger than $2L$, and construction of the $W$ matrix
leads to truncation and loss of unitarity. A consequence of this is the appearance of unphysical zero modes. This is
clearly seen in the lower left-most panel of Fig.~\ref{fig7} which shows three zero modes, while
the upper panel (free case) shows only two zero modes. Thus the truncation for the interacting case has lead to the appearance of
an additional spurious zero mode.

To understand why spurious zero modes can appear on
truncation, let us consider a $W$ with a simple $3\times 3$ structure. The $W$ have the property that
the lower sub-diagonal (c.f. Eq.~\eqref{subd})
is the strongest as it measures the part of the operator that explores new regions of the Krylov subspace. For an ergodic
system, it is natural that this element will be largest.
However, an artificial truncation will cause a $W$ of the form
\begin{align}
  W = \begin{pmatrix} 0 & 0 & 1 \\ 1 &0&0 \\0 & 1&0\end{pmatrix}
\end{align}
to become
\begin{align}
  W_{\rm trunc} =  \begin{pmatrix} 0 & 0 & 0 \\ 1 &0&0 \\0 & 1&0\end{pmatrix}
\end{align}
Now $W_{\rm trunc}$ has a null vector
\begin{align}
  W_{\rm trunc}\begin{pmatrix}0\\0\\ 1 \end{pmatrix} = 0
\end{align}
This null vector is not shared by the untruncated $W$. Thus quite generally,
the fact that the lower sub-diagonal is dominant in $W$ can lead to the appearance of null
vectors when the $W$ matrix is truncated, where the null vectors have a large weight at the lower end.

We now discuss the (A)SPMs.
The $\pm \pi$ modes are clearly visible on the top panel, and they are found to persist in the
presence of interactions, although the edge modes are not so well separated in energy from
the bulk states when interactions are present. This is expected as we now have ASPMs which will now decay into the bulk.
Sometimes the mode at $-\pi$ can appear at $\pi$ as seen in the lower fourth
panel. This is not a spurious effect
because $-\pi$ and $\pi$ states are degenerate states due to the periodicity of the spectrum.
However the truncation can lead to spurious effects such as the disappearance of one of the $\pi$ modes as can be seen in the
lower second panel.

\begin{figure*}
  \includegraphics[width = .99\textwidth]{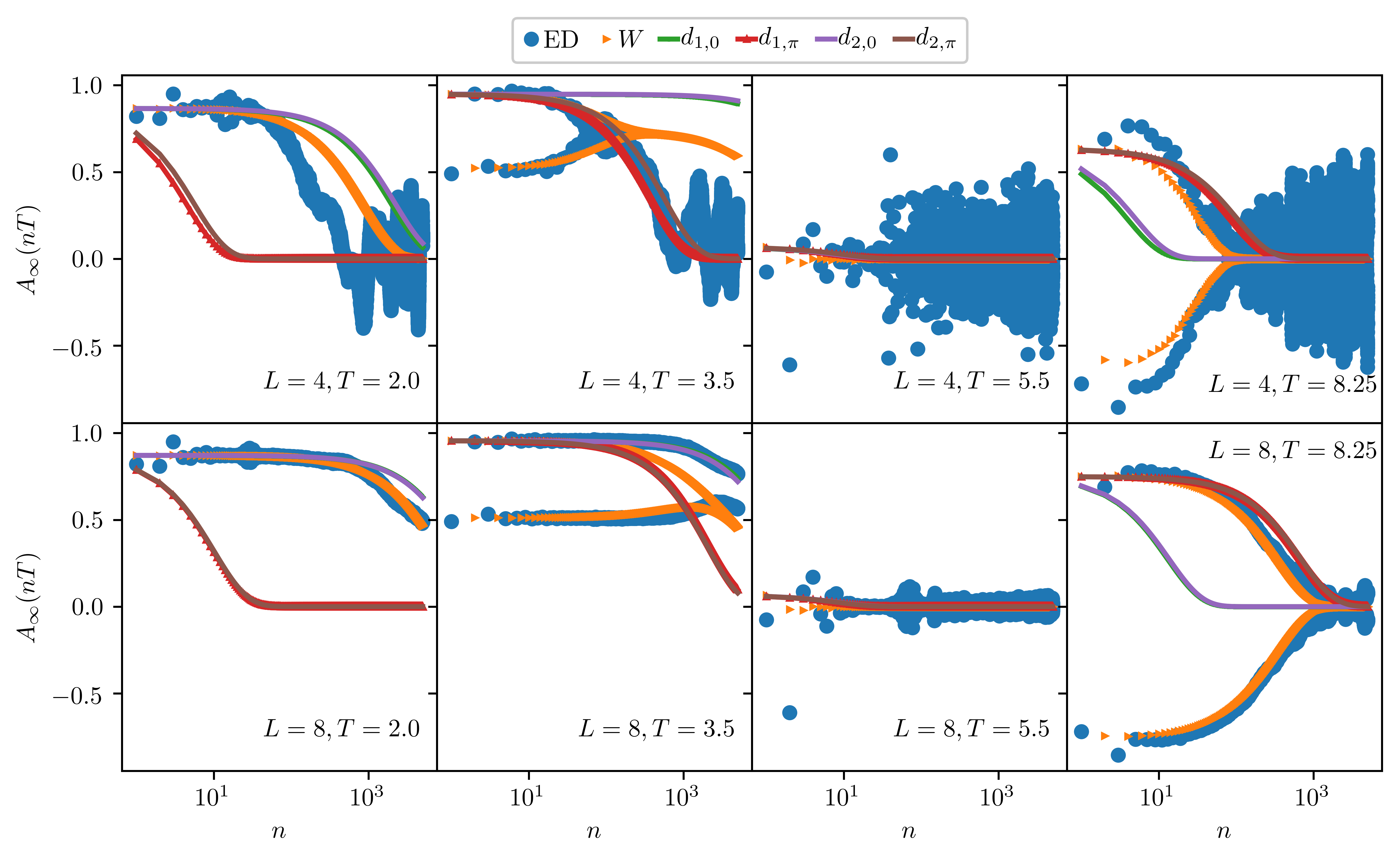}
  \caption{Time evolution of approximate edge states of $W$ (orange) for an
    interacting example ($J_z=0.05$) and compared to $A_{\infty}$ obtained from ED (blue).
    The top row is for a very small system size of $L = 4$ and the
    bottom row is for $L = 8$.  The four columns compare four
    different phases. These are from left to right: ASZM phase ($T=2.0$), ASZM-ASPM phase ($T=3.5$), trivial phase ($T=5.5$), and ASPM
    phase ($T=8.25$). $W$ is truncated to  a 18 $\times$ 18 matrix, and the time-evolution generated by it (orange) is
    according to Eq.~\eqref{ainfWN2a}.
    The data labeled by $d_{1,\lambda}, d_{2,\lambda}$ with $\lambda=0,\pi$
    are two different approximations corresponding to Eq.~\eqref{d1def} for $d_{1,\lambda}$ and Eq.~\eqref{d2def} for $d_{2,\lambda}$.}
  \label{fig8}
\end{figure*}

We now turn to the computation of $A_{\infty}$ and extracting the lifetime.
Fig.~\ref{fig8} shows the autocorrelation functions for $g=0.3$, $J_z=0.05$ and for
the same set of $T$ as Fig.~\ref{fig7}. The top row is for $L=4$ while the bottom row
is for $L=8$. The blue dots are $A_{\infty}$ from ED. The orange data are the $A_{\infty}$
obtained from the $W$ matrix,
with the $W$ matrix truncated to $N=18$. We have checked that the truncation at $N=18$ does not influence the results
for the system sizes chosen ($L=4,8$).

The precise quantity being
plotted in orange (labeled $W$) is Eq.~\eqref{ainfWN}, but with some approximations
made to it. In particular,
\begin{align}
    A_{\infty}(n T) &= {\rm Tr} \biggl[W^n |1\rangle \langle 1\biggr] \nonumber\\
    &\approx {\rm Tr} \biggl[W^n\biggl(|\psi_{0}\rangle \langle \psi_{0}| +
      |\psi_\pi \rangle \langle \psi_\pi |\biggr) |1\rangle \langle 1\biggr] \label{ainfWN2} \\
    &=\sum_{\lambda = 0,\pi}\psi_{\lambda,1}\langle 1| W^n  |\psi_{\lambda}\rangle,\,\,
    \psi_{\lambda,1}=\langle \psi_{\lambda} |1\rangle \label{ainfWN2a}.
  \end{align}
Above $|1\rangle$ is a state which is completely localized on the first site of the Krylov subspace.
The approximation made in the second line involves replacing the complete set of states of $W$ by only its edge modes
$|\psi_{\lambda=0,\pi}\rangle$. Thus we are dropping all the bulk modes of $W$.
Also note that, $\psi_{\lambda,1}$ is the normalized amplitude of the edge mode
on the first site.

The results for $A_{\infty}$  in Fig.~\ref{fig8} are compared to two different approximations. One
is the data labeled as $d_{1,\lambda=0,\pi}$ where
\begin{align}
  d_{1,\lambda}(n T) &= |\psi_{\lambda,1}|^2e^{-\gamma_{\lambda} n}\nonumber\\
&=  |\psi_{\lambda,1}|^2|\langle \psi_{\lambda}|W|\psi_{\lambda}\rangle|^{n}, \,\, \lambda=0,\pi.\label{d1def}
\end{align}
The above expression uses the definition of the decay-rate in Eq.~\eqref{eqlt1} accounting for
the amplitude $\psi_{\lambda, 1}$ of finding the particle on the first site, before the time-evolution.
Since $|\psi_{0,\pi}\rangle$ edge modes
are only approximate edge modes, Eq.~\eqref{d1def} does lead to a decay.
The data set labeled by $d_{2,\lambda=0,\pi}$ corresponds to approximating
$\gamma_{0,\pi}$ by Eq.~\eqref{eqlifetime3}, thus
\begin{align}
d_{2,\lambda}(n T) = |\psi_{\lambda,1}|^2e^{-|\psi_{\lambda, N}|^2n}, \,\, \lambda=0,\pi,\label{d2def}
\end{align}
where $\psi_{\lambda, N}$ is the normalized amplitude of the $\lambda=0,\pi$ approximate edge mode at
the last site $N$.

We find that $W$, despite its truncation, agrees very well with ED. In addition, the two approximations to $W$
given by Eq.~\eqref{d1def} and Eq.~\eqref{d2def} also agree well as far as capturing the decay rates are concerned.
The approximations appear to not work quite so well for the second panel corresponding to
the ASZM-ASPM phase ($T=3.5$). This may be because when both almost strong modes are present, as they decay they influence each other
through the bulk states in a way that our simple approximation has not accounted for.
However, the agreement does improve with increasing system size (compare upper
and lower panels for $T=3.5$). In particular, when both ASZM and ASPM are present,
the edge structure is more complex and has a $2\times 2$ structure at each end (see Section \ref{Arnoldi-free}).
This can make the effective system size of $L=4$ (upper panels) look effectively shorter for an ASZM-ASPM phase
as compared to an ASPM or an ASZM phase.

\section{Conclusions}\label{Conclu}

Almost strong edge modes are quasi-stable edge modes that have unusually long lifetimes such that they
can coexist with a metallic bulk for many drive cycles.
These edge modes include both zero modes as well as $\pi$ modes, where the latter show
period doubled dynamics.
It is notoriously hard to develop
analytic methods for interacting and driven systems in general, and determining lifetimes of quasi-stable modes in
particular.
However in this paper, building on previous work \cite{Yates20,Yates20a,Yates21a},  we showed a promising route which involves
mapping the dynamics of the edge mode operator to single particle dynamics in Krylov subspace.
While the detailed modeling of the Krylov
subspace, such as the precise values of the hopping parameters, still requires the same computational costs as ED,
yet when the operator of interest has some universal features,
the Krylov subspace can have some general properties. For example, the Krylov subspace of maximally chaotic systems
have certain universal features much discussed in the literature \cite{Altman19,Gorsky19,Sinha19,Avdoshkin19}.

In this paper we showed that when the operators are strong modes or almost strong modes,
they appear as stable or quasi-stable edge modes of Krylov subspaces with topological properties. For the examples studied in
this paper, the Krylov subspaces
are given by generalized SSH models with long range and spatially inhomogeneous hopping.
Exploiting these topological structures
can lead to better understanding of the long lifetimes of the almost strong modes.

Most studies on Krylov subspace dynamics use the Lanczos method where the generator of
dynamics is a static Hamiltonian \cite{Recbook,Altman19,Gorsky19,Sinha19,Avdoshkin19}.
In studying Floquet systems, as we do in this paper, one has to adapt the Krylov subspace because the generator of dynamics
is a unitary operator rather than a Hermitian operator.
This lead us to derive the Krylov subspace using an alternate approach, known as the
the Arnoldi iteration \cite{Arnoldi51}. Topological features of the resulting chain were highlighted. A compact expression
(c.f. Eq.~\eqref{d2def}) for the lifetime of the edge modes was derived and compared with ED.
In particular, we showed that the decay rate is simply determined
by the probability of finding the particle on the last site $N$ of the $W$ matrix that generates the time-evolution
in this Krylov subspace. This observation opens up the possibility of developing
analytic tools for calculating the decay rate, such as by employing the WKB approximation.

Our methods can be generalized to study other kinds of slow dynamics such as dynamics of scar
states \cite{Bernevig20,Papic18,Lukin19,Motrunich19}. It is also promising to
perform a topological classification of the Krylov subspaces of edge modes of strongly interacting topological
insulators, both static and Floquet \cite{Fidkow10,Potter16, Else16b}.

{\sl Acknowledgements:}
The authors thank Sasha Abanov for helpful discussions.
This work was supported by the US Department of Energy,
Office of Science, Basic Energy Sciences, under Award No.~DE-SC0010821.


%

\end{document}